\documentclass[letterpaper,titlepage,11pt]{article}
\usepackage[colorlinks=true,urlcolor=blue,citecolor=magenta]{hyperref}
\usepackage{amssymb,amsmath,amsfonts}
\usepackage{color}
\usepackage{graphicx}
\usepackage{array}

\def\clock{{\count0=\time
           \divide\count0 60
           \ifnum\count0<10 0\fi\the\count0
           \multiply\count0 -60 \advance\count0 \time
           :\ifnum\count0<10 0\fi \the\count0
         }}
\newcommand{\timestamp}{{\small\vbox{\hbox{\tt\jobname.tex}
\hbox{\the\day/\the\month/\the\year, \clock}}}}

\newcommand{\beq}{\begin{equation}}
\newcommand{\eeq}{\end{equation}}

\let\oldsqrt\sqrt
\def\sqrt{\mathpalette\DHLhksqrt}
\def\DHLhksqrt#1#2{%
\setbox0=\hbox{$#1\oldsqrt{#2\,}$}\dimen0=\ht0
\advance\dimen0-0.2\ht0
\setbox2=\hbox{\vrule height\ht0 depth -\dimen0}%
{\box0\lower0.4pt\box2}}

\newcommand{\bv}{\boldsymbol{\zeta}}
\numberwithin{equation}{section}

\begin{document}

\hypersetup{pageanchor=false}
\begin{titlepage}
 \vskip 1.8 cm

\centerline{\Huge \bf Electroelasticity of Charged Black Branes}
\vskip 2cm

\centerline{\large {\bf Jay Armas$^{1}$, Jakob Gath$^{2}$ and Niels A. Obers$^{2}$}}

\vskip 1.0cm

\begin{center}
\sl $^{1}$ Albert Einstein Center for Fundamental Physics \\
\sl Institute for Theoretical Physics, University of Bern \\
\sl  Siddlerstrasse 5, 3012-Bern, Switzerland
\end{center}
\vskip 0.4cm

\begin{center}
\sl $^2$ The Niels Bohr Institute, Copenhagen University  \\
\sl  Blegdamsvej 17, DK-2100 Copenhagen \O , Denmark
\end{center}

\centerline{\small\tt jay@itp.unibe.ch, gath@nbi.dk, obers@nbi.dk}

\vskip 1.3cm \centerline{\bf Abstract} \vskip 0.2cm \noindent

We present the first-order corrected dynamics of fluid branes carrying higher-form charge by obtaining the general form of 
their equations of motion to pole-dipole order.  
Assuming linear response theory, we characterize the corresponding effective theory of stationary bent charged (an)isotropic
fluid branes in terms of two sets of response coefficients, the Young modulus and the piezoelectric moduli. 
We subsequently find large classes of examples in gravity of this 
effective theory, by constructing stationary strained charged black brane solutions to first order in a derivative expansion. 
Using solution generating techniques
and bent neutral black branes as a seed solution, we obtain a class of charged black brane geometries carrying smeared 
Maxwell charge in Einstein-Maxwell-dilaton gravity. 
In the specific case of ten-dimensional space-time we furthermore use T-duality to generate bent black branes with higher-form charge, including smeared D-branes of type II string theory. 
By subsequently measuring the bending moment and the electric dipole moment which these geometries acquire due to the strain, 
we uncover that their form is captured by classical electroelasticity theory. In particular, we find that the Young modulus and the piezoelectric moduli of our strained charged black brane solutions are parameterized by a total of 4 response coefficients, both for the isotropic
as well as anisotropic cases.

\end{titlepage}
\hypersetup{pageanchor=true}


\tableofcontents


\section{Introduction} \label{intro}
Long-wavelength perturbations of black branes have been useful for the construction of new black hole solutions in higher dimensions,
 as well as for understanding finite temperature properties of strongly coupled quantum field theories by means of holographic dualities. In this long-wavelength regime, black branes behave much like any other type of continuous media whose dynamics is governed by specific effective theories. Two types of deformations to black branes have been considered in the literature: time-(in)dependent fluctuations along the boundary/worldvolume directions \cite{Bhattacharyya:2008jc,Bhattacharyya:2008mz,Erdmenger:2008rm,Banerjee:2008th,Camps:2010br,Gath:2013qya,Emparan:2013ila} and stationary perturbations along directions transvere to the worldvolume \cite{Emparan:2007wm,Caldarelli:2008pz,Emparan:2009vd,Armas:2011uf,Camps:2012hw,Armas:2012ac}. The former are characterized by an effective theory of viscous fluid flows \cite{Bhattacharyya:2008jc} while the latter are characterized by an effective theory of thin elastic branes \cite{Emparan:2009cs, Emparan:2009at,Armas:2012jg,Armas:2013hsa}. Both of these descriptions are unified in a general framework of fluids living on dynamical surfaces (fluid branes), which, when applied to black branes, is known collectively as the blackfold approach \cite{Emparan:2009cs, Emparan:2009at}. In this framework, Ref.~\cite{Armas:2013hsa} recently provided, in the probe brane approximation, the 
leading order corrections to the effective action for stationary neutral fluid branes.  

This paper is concerned with stationary elastic perturbations along transverse directions to the worldvolume of charged black branes. This type of deformation is achieved by breaking the symmetries of the transverse space to the brane worldvolume directions in the same way that the circular cross-section of a rod is deformed when it is bent. Such perturbations have been studied in \cite{Emparan:2007wm,Caldarelli:2008pz,Emparan:2009vd,Armas:2011uf,Camps:2012hw} for neutral black branes and in \cite{Armas:2012ac} for charged asymptotically flat dilatonic black strings bent into a circular shape. In these cases, to first order in the derivative expansion, the metric acquires a bending moment while, in the case of charged branes, the gauge field acquires an electric dipole moment which encode the brane response to applied strains. The reader should be reminded of a consequence of placing a fluid on a dynamical surface (submanifold) embedded in a background space-time: if the surface is deformed along transverse directions, the induced metric changes and that change is the measure of the strain \cite{Armas:2011uf, Armas:2012jg}.

According to the classical theory of elasticity, the bending moment encodes the response coefficients of the material to applied strains \cite{Landau:1959te}. For a generic material these coefficients are a set of elastic moduli that are described by a tensor structure with the name of Young modulus. For the case of neutral black branes these have been measured in 
\cite{Armas:2011uf,Camps:2012hw} and have been recently classified using the general framework of \cite{Armas:2013hsa}. If the material is electrically charged, according to the theory of electroelasticity, the gauge field will develop an electric dipole moment whose strength is proportional to a set of piezoelectric moduli \cite{bardzokas2007mathematical}. This effect was first measured in \cite{Armas:2012ac} for asymptotically flat charged dilatonic black strings in Einstein-Maxwell-dilaton (EMD) gravity\footnote{For another occurrence of the piezoelectric effect in the context of superfluids see \cite{Erdmenger:2012zu}.}.
In order to compute the response coefficients of black branes with higher-form charges, we first need to derive the first order corrected dynamics of charged fluid branes by obtaining the general form of  their equations of motion to pole-dipole order.  This generalizes the analysis done in \cite{Armas:2011uf} for neutral banes and in \cite{Armas:2012ac} for branes with Maxwell charge.
In particular, our analysis will include $p$-branes carrying $q$-brane charge where $p>q$ and $q\ne0$, which are characterized by a stress-energy tensor of an anisotropic fluid \cite{Grignani:2010xm,Caldarelli:2010xz,Emparan:2011hg}. Therefore, the subsequent measurement of response coefficients in the corresponding bent charged black brane geometries involves 
measuring for the first time the Young modulus and piezoelectric moduli of anisotropic fluid branes.

The usefulness of bending charged branes can be understood if one remembers the origin of the blackfold approach, namely, the construction of new black hole solutions by wrapping black branes along a submanifold with the desired topology \cite{Emparan:2009cs,Emparan:2009at,Emparan:2009vd}. To better understand this imagine playing the following game where one has to construct a circular geometry out of twenty pieces of straw. To leading order in this construction one places each of the twenty straws tangent to a circular line. To first order, one should slightly bend each of the individual straws into a circular shape with the same radius of curvature as the radius of the circle along which the straws are placed and so on to higher orders. Increasing the number of straws from twenty to infinity so that the geometry becomes continuous and replacing each individual straw by a black brane allows one to construct many new black hole solutions in a perturbative manner \cite{Emparan:2007wm,Caldarelli:2008pz,Emparan:2009cs,Emparan:2009vd,Armas:2010hz,Caldarelli:2010xz,Emparan:2011hg} and requires solving the blackfold equations \cite{Emparan:2009at,Armas:2011uf,Camps:2012hw,Armas:2013hsa}. To leading order, information about the response coefficients due to bending is not required but to higher orders they constitute the necessary data that serve as input into the effective theory \cite{Armas:2013hsa}. Another valuable outcome of the blackfold approach is the connection between its effective theory and that of systems studied in theoretical biology \cite{Armas:2013hsa} as well as improved effective actions for QCD
\cite{Polyakov:1986cs,Kleinert:1986bk}. Therefore, uncovering the structure of the response coefficients for blackfold geometries 
provides novel insights into the general structure of the effective theory for these systems. 

We find in this work large classes of examples in gravity of the electroelastic phenomena suggested by our general analysis
of the equations of motion at pole-dipole order of charged fluid branes and the associated linear response theory. 
In particular, we use standard solution generating techniques and the bent neutral black branes of  \cite{Emparan:2007wm,Camps:2012hw} as seed solutions, to construct stationary strained charged black brane geometries to first order in a derivative expansion.
In this way we first obtain a general class of bent charged black brane geometries carrying smeared Maxwell (0-brane) charge in Einstein-Maxwell-dilaton gravity with Kaluza-Klein coupling constant, which were first considered in \cite{Armas:2012ac}. The corresponding effective theory describing the perturbed solution is that of an isotropic fluid brane which has been subject to pure bending. 
In the specific case of ten-dimensional space-time we furthermore use T-duality to generate bent black branes charged under higher-form fields. This includes type II D$q$-branes smeared in $(p-q)$-directions, which are thus described by
the theory of anisotropic $p$-branes carrying $q$-brane charge. 
By subsequently measuring the bending moment and the electric dipole moment which these geometries acquire due to the strain, 
we uncover that their form is captured by classical electroelasticity theory and we can determine
their Young modulus and piezoelectric moduli. For the class of bent charged black brane solutions obtained in this paper,
these are parameterized by a total of 4 response coefficients, both for the isotropic as well as anisotropic cases. 
These measurements constitute the first step in obtaining higher order corrections to the charged stationary black holes found in \cite{Caldarelli:2010xz,Emparan:2011hg}.

The outline of this paper is as follows. In Sec.~\ref{sec:dynamics} we derive the equations of motion for $p$-branes carrying  $q$-brane charge along their worldvolume using the methods of \cite{Vasilic:2007wp,Armas:2011uf}. We first present the case $q=0$, and subsequently treat the simplest anistropic case $q=1$ in detail, while presenting the main results for the case $q>1$. 
Sec.~\ref{sec:physical}  gives a physical interpretation for the different structures appearing in the multipole expansion of the electric current, and at the same time reviews the corresponding results relevant to the stress-energy tensor.
For the purpose of this paper, the most important quantity in the expansion of the electric current is the electric dipole moment, but we also 
briefly comment on the magnetic dipole moment, which enters the description when considering spinning branes.  
We also  define and classify the response coefficients encoding the response of charged fluid branes due to electroelastic deformations according to the expectation from classical electro-elastodynamics. In Sec.~\ref{sec:measurement} we outline the measurement procedure for these response coefficients for charged black branes in gravity. In particular, 
we provide explicit expressions for these coefficients for branes carrying $0$-brane charge in EMD gravity and for branes carrying $q$-brane charge in ten-dimensional type II string theory generalizing the analysis of \cite{Armas:2012ac}. 
In Sec.~\ref{sec:discussion} we comment on open issues and future work. We also provide three appendices.  In App.~\ref{app:eom} we present details on the equations of motion for the case of branes carrying string charge. 
In App.~\ref{app:neutral} we review the case of elastic perturbations of neutral black branes \cite{Emparan:2007wm, Camps:2012hw} and provide the map between the conventions used here and others used previously in the literature. 
Finally, in App.~\ref{app:metrics} we give an outline of the solution generating techniques as well as the explicit form of the geometries constructed in this paper.


\section{Dynamics of charged pole-dipole branes} \label{sec:dynamics}
In this section we obtain the equations of motion for pole-dipole branes carrying Maxwell charge ($q=0$) by solving conservation equations for the effective stress-energy tensor and effective current that characterize the branes. We subsequently provide the equations of motion for branes carrying string charge ($q=1$) as well as for branes charged under higher-form fields. Further details of the derivation of these equations are given in App.~\ref{app:eom}.

\subsection{Effective stress-energy tensor and effective current}
To probe the effect of bending an object requires giving the object a finite thickness. Physically, this is because bending induces a varying concentration of matter along transverse directions to the brane worldvolume resulting in a non-trivial bending moment \cite{Armas:2011uf}. If the brane is infinitely thin, bending effects cannot be taken into account. In order to include finite thickness effects in the brane dynamics one performs a multipole expansion of the stress-energy tensor in the manner \cite{Vasilic:2007wp}
\beq \label{stpd}
T^{\mu\nu}(x^{\alpha})=\int_{\mathcal{W}_{p+1}}\!\!\!\!\!d^{p+1}\sigma\sqrt{-\gamma}\left[T^{\mu\nu}_{(0)}(\sigma^{a})\frac{\delta^{D}(x^{\alpha}-X^{\alpha}(\sigma^{a}))}{\sqrt{-g}}-\nabla_{\rho}\left(T_{(1)}^{\mu\nu\rho}(\sigma^a)\frac{\delta^{D}(x^{\alpha}-X^{\alpha}(\sigma^{a}))}{\sqrt{-g}}\right)+...\right]~~\!\!.
\eeq
A few remarks about our conventions are now in place. We consider a $(p+1)$-dimensional worldvolume embedded in a $D=n+p+3$ space-time with coordinates $x^{\alpha}$ and metric $g_{\mu\nu}$, where the greek indices $\mu\nu,...$ label space-time directions. The worldvolume $\mathcal{W}_{p+1}$ is parameterized by a set of coordinates $\sigma^{a}$ and inherits an induced metric $\gamma_{ab}=g_{\mu\nu}u_{a}^{\mu}u_{b}^{\nu}$ with $u_{a}^{\mu}\equiv \partial_{a}X^{\mu}$. The location of $\mathcal{W}_{p+1}$ in space-time is given by the set of mapping functions $X^{\mu}(\sigma^a)$. 

The stress-energy tensor \eqref{stpd} is characterized by two structures: $T^{\mu\nu}_{(0)}$ is a monopole source of stress-energy while $T_{(1)}^{\mu\nu\rho}$ encodes the dipole (finite thickness) effects. To each of these structures one associates an order parameter $\tilde\varepsilon$ such that $T^{\mu\nu}_{(0)}=\mathcal{O}(1)$ and $T_{(1)}^{\mu\nu\rho}=\mathcal{O}(\tilde\varepsilon)$. Typically, for branes of thickness $r_0$ bent over a submanifold of  characteristic curvature radius $R$, the parameter $\tilde\varepsilon$ has the form $\tilde\varepsilon=r_0/R$. If the expansion \eqref{stpd} is truncated to $\mathcal{O}(\tilde\varepsilon)$~, the stress-energy tensor is said to be expanded to pole-dipole order. When only hydrodynamic corrections are considered, the stress-energy tensor is localized on the surface described by $X^{\mu}(\sigma^a)$ due to the delta-function in \eqref{stpd}, since in this case $T_{(1)}^{\mu\nu\rho}=0$ while $T^{\mu\nu}_{(0)}$ receives viscous corrections order-by-order in a derivative expansion. However, when elastic perturbations are considered the brane acquires a bending moment which is encoded in $T_{(1)}^{\mu\nu\rho}$. In this case, there exists an ambiguity in the position of the worldvolume surface within a finite region of thickness $r_0$, which is parametrized by the `extra symmetry 2' acting on $T^{\mu\nu}_{(0)}$ and $T_{(1)}^{\mu\nu\rho}$ under a $\mathcal{O}(\tilde\varepsilon)$ displacement of the worldvolume location $X^{\alpha}(\sigma^{a})\to X^\alpha(\sigma^a)+\tilde\varepsilon^{\alpha}(\sigma^a)$ \cite{Vasilic:2007wp}. 

The equations of motion for an object with a stress-energy tensor of the type \eqref{stpd}, assuming the absence of external forces and ignoring backreaction, follow from the conservation equation
\beq
\nabla_{\nu}T^{\nu\mu}=0~~.
\eeq
In order to write the equations of motion in a way adapted to the cases considered in this paper, we decompose $T_{(1)}^{\mu\nu\rho}$ into tangential and orthogonal parts with the help of the orthogonal projector ${\perp^{\mu}}_{\nu}={\delta^{\mu}}_{\nu}-u^{\mu}_{a}u^{a}_{\nu}$ such that
\beq \label{decomp_T}
T_{(1)}^{\mu\nu\rho}={u_{b}}^{(\mu}j^{(b)\nu)\rho}+u^{\mu}_{a}u^{\nu}_{b}d^{ab\rho}+u^{\rho}_{a}T^{\mu\nu a}_{(1)}~~.
\eeq
Here the parenthesis in the index $b$ appearing in the first structure in \eqref{decomp_T} indicate that this index is insensitive to the symmetrization which is done only over the space-time indices $\mu,\nu$. Moreover, $j^{b\nu\rho}$ are the components responsible for giving transverse motion (spin) to the brane and have been considered by Papapetrou when deriving the equations of motion for spinning point particles \cite{Papapetrou:1951pa}. These have the properties $j^{b\nu\rho}=j^{b[\nu\rho]}$ and $u_{\nu}^{a}j^{b \nu\rho}=0$. The components $d^{ab\rho}$ have the properties $d^{ab\rho}=d^{(ab)\rho}$ and $u^{c}_{\rho}d^{ab\rho}=0$ and encode the bending moment of the brane. In the point particle case these components can be gauged away using the `extra symmetry 2' \cite{Armas:2011uf} but not for the cases $p>0$. The components $T^{\mu\nu a}_{(1)}$ can be gauged away everywhere on the worldvolume using the `extra symmetry 1' and can be set to zero at the boundary in the absence of additional boundary sources \cite{Vasilic:2007wp}. As we are only interested in bending corrections, i.e. $j^{b\nu\rho}=0$, we can write the equations of motion as \cite{Armas:2011uf, Armas:2013hsa}
\beq\label{st11}
\nabla_{a}\hat{T}^{ab}+u^{b}_{\mu}\nabla_{a}\nabla_{c}d^{ac\mu}=d^{ac\mu}{R^{b}}_{ac\mu}~~,
\eeq
\beq \label{st12}
\hat T^{ab}{K_{ab}}^{\rho}+{\perp^{\rho}}_{\mu}\nabla_{a}\nabla_{b}d^{ab\mu}=d^{ab\mu}{R^{\rho}}_{ab\mu}~~.
\eeq
Here, ${R^{\rho}}_{\nu\lambda\mu}$ is the Riemann curvature tensor of the background space-time, ${K_{ab}}^{\rho}\equiv\nabla_{a}u^{\rho}_{b}$ is the extrinsic curvature tensor of $\mathcal{W}_{p+1}$\footnote{Here we have introduced the worldvolume covariant derivative $\nabla_{a}\equiv u^{\rho}_{a}\nabla_{\rho}$ which acts on a generic space-time tensor $V^{c\mu}$ as $\nabla_{a}V^{c\mu}=\partial_{a}V^{c\mu}+{\gamma_{ab}}^{c}V^{b\mu}+{\Gamma^{\mu}}_{\lambda\rho}u^{\lambda}_{a}V^{c\rho}$, where ${\gamma_{ab}}^{c}$ are the Christoffel symbols associated with $\gamma_{ab}$ and ${\Gamma^{\mu}}_{\lambda\rho}$ are the Christoffel symbols associated with $g_{\mu\nu}$.} and $\hat{T}^{ab}=T^{ab}_{(0)}+2d^{(ac\mu}{K^{b)}}_{c\mu}$. If finite thickness effects are absent, $d^{ab\mu}=0$, one recovers the equations of motion derived by Carter \cite{Carter:2000wv} and if one further takes $T^{ab}_{(0)}$ to be of the perfect fluid form with energy density and pressure of a black brane, these equations are the leading order blackfold equations \cite{Emparan:2009cs, Emparan:2009at}. Eqs.~\eqref{st11}-\eqref{st12} are relativistic generalizations of the equations of motion of thin elastic branes \cite{Armas:2013hsa} and must be supplemented with the integrability condition $d^{ab[\mu}{K_{ab}}^{\rho]}=0$ and boundary conditions
\beq \label{st13}
d^{ab\rho}\eta_a\eta_b|_{\mathcal{W}_{p+1}}=0~~,~~\left(\hat{T}^{ab}u_{b}^{\mu}-d^{ac\rho}{K^{b}}_{c\rho}{u_{b}}^{\mu}+{\perp^{\mu}}_{\rho}\nabla_{b}d^{ab\rho}\right)\eta_a|_{\mathcal{W}_{p+1}}=0~~,
\eeq
where $\eta_a$ is a unit normal vector to the brane boundary. 

Eqs.~\eqref{st11}-\eqref{st13} are also valid for charged (dilatonic) branes as long as couplings to external background fields are absent \cite{Caldarelli:2010xz, Emparan:2011hg, Grignani:2010xm}. However, if the brane is charged under a $(q+1)$ gauge field, it is also characterized by a total anti-symmetric current tensor $J^{\mu_{1}...\mu_{q+1}}$ which can also be expanded in a Dirac-delta series as \cite{Armas:2012ac}
\beq \label{cpd}
J^{\mu_{1}...\mu_{q+1}}(x^{\alpha})=\int_{\mathcal{W}_{p+1}}\!\!\!\!\!d^{p+1}\sigma\sqrt{-\gamma}\left[J^{\mu_1...\mu_{q+1}}_{(0)}\frac{\delta^{D}(x^{\alpha}-X^{\alpha})}{\sqrt{-g}}-\nabla_{\rho}\left(J_{(1)}^{\mu_1...\mu_{q+1}\rho}\frac{\delta^{D}(x^{\alpha}-X^{\alpha})}{\sqrt{-g}}\right)+...\right]~~\!\!,
\eeq
where we have omitted the explicit dependence of $J^{\mu_1...\mu_{q+1}}_{(0)},J_{(1)}^{\mu_1...\mu_{q+1}\rho}$ and $X^{\alpha}$ on the worldvolume coordinates $\sigma^a$. As in the case of the stress-energy tensor \eqref{stpd} the structure $J^{\mu_1...\mu_{q+1}}_{(0)}$ is a monopole source of a charged $q$-brane current while the structure $J_{(1)}^{\mu_1...\mu_{q+1}\rho}$ encodes the finite thickness effects, including the electric dipole moment of the brane. Moreover, the structures involved in \eqref{cpd} follow the same hierarchy as in the case of the \eqref{stpd}, i.e., $J^{\mu_1...\mu_{q+1}}_{(0)}=\mathcal{O}(1)$ and $J_{(1)}^{\mu_1...\mu_{q+1}\rho}=\mathcal{O}(\tilde\varepsilon)$. The equations of motion for the current \eqref{cpd} follow from the conservation equation
\beq \label{ceom}
\nabla_{\mu_1}J^{\mu_1...\mu_{q+1}}=0~~.
\eeq
The equations of motion that follow from here have been previous derived by Dixon and Souriou in \cite{Dixon:1974, Souriou:1974} for charged point particles ($q=0$)\footnote{See Ref.~\cite{Costa:2012cy} for a recent review, including a treatment of the case when external forces are present.} in a different way than the one presented in \cite{Armas:2012ac}. In the next few sections we will derive these equations in detail for $p$-branes carrying $0$-brane charge and then present the results for generic $p$-branes carrying $q$-brane charge. Further details about their derivation are given in App.~\ref{app:eom}.


\subsection{Equations of motion for branes carrying Maxwell charge} \label{sec:stringcharge}
Pole-dipole $p$-branes carrying Maxwell charge ($q=0$) are characterized by a current $J^{\mu}$ of the form \eqref{cpd}. In order to solve Eq.~\eqref{ceom} we introduce an arbitrary scalar function $f(x^\alpha)$ of compact support and integrate \eqref{ceom} over the entire space-time following the method outlined in \cite{Vasilic:2007wp} applied to the stress-energy tensor \eqref{stpd}
\beq \label{intceom}
\int d^{D}x\sqrt{-g}f(x^{\alpha})\nabla_{\mu}J^{\mu}=0~.
\eeq
In order to make further progress one decomposes the derivatives of $f(x^{\alpha})$ in parallel and orthogonal components to the worldvolume such that
\beq
\nabla_{\mu}f=f_{\mu}^{\perp}+u^{a}_{\mu}\nabla_{a}f~~,~~\nabla_{\nu}\nabla_{\mu}f=f_{\mu\nu}^{\perp}+2f^{\perp}_{(\mu a}u^{a}_{\nu)}+f_{ab}u^{a}_{\mu}u^{b}_{\nu}~~.
\eeq
Here the label $\perp$ on a tensor indicates that all of its space-time indices are transverse, for example, $u^{\mu}_{a}f_{\mu}^{\perp}=0$. Explicit computation of the functions involved allows one to deduce
\beq
f^{\perp}_{\mu a}={\perp^{\lambda}}_{\mu}\nabla_{a}f^{\perp}_{\lambda}+\left(\nabla_{a}u^{b}_{\mu}\right)\nabla_{b}f~~,~~f_{ab}=\nabla_{(a}\nabla_{b)}f-f_{\mu}^{\perp}\nabla_{b}u_{a}^{\mu}~~.
\eeq
This tells us that the only independent components on the worldvolume surface $x^{\alpha}=X^{\alpha}(\sigma^a)$ are $f^{\perp}_{\mu\nu}$, $f^{\perp}_{\mu}$ and $f$. Using this and performing a series of partial integrations when introducing \eqref{cpd} into \eqref{intceom} leads to an equation with the following structure
\beq \label{Zs}
\int_{\mathcal{W}_{p+1}}\sqrt{-\gamma}\left[Z^{\mu\nu}f_{\mu\nu}^{\perp}+Z^{\mu}f_{\mu}^{\perp}+Zf+\nabla_{a}\left(Z^{\mu a}f_{\mu}^{\perp}+Z^{ab}\nabla_{b}f+Z^{a}f\right)\right]=0~~. 
\eeq
Requiring the above equation to vanish for each of the arbitrary independent components on the worldvolume $f^{\perp}_{\mu\nu}$, $f^{\perp}_{\mu}$ and $f$ results in the equations
\beq \label{eom10}
{\perp^{\lambda}}_{\mu}{\perp^{\rho}}_{\nu}J_{(1)}^{(\mu\nu)}=0~~,~~{\perp^{\lambda}}_{\mu}\left[J_{(0)}^{\mu}-\nabla_{a}\left(2{\perp^{\mu}}_{\lambda}u^{a}_{\rho}J^{(\lambda\rho)}_{(1)}+u^{a}_{\nu}u^{b}_{\rho}u^{\mu}_{b}J^{(\nu\rho)}_{(1)}\right)\right]=0~~,
\eeq
\beq\label{eom30}
\nabla_{a}\left(J^{\mu}_{(0)}u^{b}_{\mu}+2J^{(\mu\nu)}_{(1)}u^{b}_{\nu}\nabla_{b}u^{a}_{\mu}-\nabla_{b}\left(J^{(\mu\nu)}_{(1)}u^{a}_{\mu}u^{b}_{\nu}\right)\right)=0~~.
\eeq
From Eq.~\eqref{Zs}, we are then left with a boundary term that vanishes by itself
\beq \label{Zb}
\int_{\partial\mathcal{W}_{p+1}}\sqrt{-h}\eta_{a}\left(Z^{\mu a}f_{\mu}^{\perp}+Z^{ab}\nabla_{b}f+Z^{a}f\right)=0~~,
\eeq
where $h$ is the determinant of the induced metric on the boundary. 

On the brane boundary, however, the components $\nabla_{a}f$ are not independent so we decompose them according to
\beq
\nabla_{a}f=\eta_{a}\nabla_{\perp}f+v_{a}^{\hat{a}}\nabla_{\hat{a}}f~~,
\eeq
where $\nabla_{\perp}\equiv\eta^{a}\nabla_{a}$, $v_{a}^{\hat{a}}$ are boundary coordinate vectors and $\nabla_{\hat{a}}$ is the boundary covariant derivative with the indices $\hat{a}$ labeling boundary directions. On the brane boundary the functions $f_{\mu}^{\perp}$, $\nabla_{\perp}f$ and $f$ are mutually independent. Requiring the terms appearing in Eq.~\eqref{Zb} proportional to these functions to vanish leads to the following boundary conditions:
\beq \label{b10}
{\perp^{\lambda}}_{\mu}J^{(\mu\nu)}_{(1)}u^{a}_{\nu}\eta_a|_{\partial\mathcal{W}_{p+1}}=0~~,~~J^{(\mu\nu)}_{(1)}u^{a}_{\mu}u^{b}_{\nu}\eta_a\eta_b|_{\partial\mathcal{W}_{p+1}}=0~~,
\eeq
\beq \label{b20}
\left[\nabla_{\hat{a}}\left(J^{(\mu\nu)}_{(1)}u^{a}_{\mu}u^{b}_{\nu}v^{\hat{a}}_{b}\eta_a\right)-\eta_b\left(J^{\mu}_{(0)}u^{b}_{\mu}+2J^{(\mu\nu)}_{(1)}u^{a}_\nu\nabla_{a}u^{b}_{\mu}-\nabla_{a}\left(J^{(\mu\nu)}_{(1)}u^{a}_{\mu}u^{b}_\nu\right)\right)\right]|_{\partial\mathcal{W}_{p+1}}=0~~.
\eeq
We now wish to solve the equations of motion \eqref{eom10}-\eqref{eom30}. To that end, we make the most general decomposition of $J^{\mu}_{(0)}$ and $J^{\mu\nu}_{(1)}$ in terms of tangential and orthogonal components such that
\beq \label{decomp0}
J^{\mu}_{(0)}=J_{(0)}^{a}u^{\mu}_{a}+J^{\mu}_{\perp(1)}~~,~~J^{\mu\nu}_{(1)}=m^{\mu\nu}+u^{\mu}_{a}p^{a\nu}+J^{\mu a}_{(1)}u^{\nu}_{a}~~,
\eeq
where $m^{\mu\nu}$ is transverse in both indices and $p^{a\rho}$ is transverse in its space-time index. The structure $J^{\mu a}_{(1)}$ is left neither parallel nor orthogonal to the worldvolume and further satisfies $J^{[ab]}_{(1)}=0$. The reason for this will become clear below. We now introduce the decomposition of $J^{\mu\nu}_{(1)}$ into the first equation in \eqref{eom10} and obtain the constraint $m^{\mu\nu}=m^{[\mu\nu]}$. Introducing both decompositions \eqref{decomp0} into the second equation in \eqref{eom10} leads to the relation
\beq \label{motion10}
J^{\mu}_{\perp(1)}={\perp^{\mu}}_{\nu}\nabla_{a}\left(p^{a\nu}+J^{\nu a}_{(1)}\right)~~.
\eeq
Further, using \eqref{decomp0} in Eq.~\eqref{eom30} leads to the equation for worldvolume current conservation 
\beq \label{motion20}
\nabla_{a}\left(\hat{J}^{a}+p^{b\mu}{K^{a}}_{b\mu}\right)=0~~,
\eeq
where $\hat{J}^{a}=J^{a}_{(0)}-u^{a}_{\mu}\nabla_{b}J^{\mu b}_{(1)}$. Note that in the case $J^{\mu\nu}_{(1)}=0$ for which the brane is infinitely thin, the equation of motion \eqref{motion20} reduces to that obtained previously in the literature using the same method \cite{Armas:2012bk}. Turning now to the boundary conditions \eqref{b10}-\eqref{b20} using \eqref{decomp0} we obtain
\beq \label{bound10}
\left(p^{a\mu}+J^{\mu a}_{\perp(1)}\right)\eta_a|_{\partial\mathcal{W}_{p+1}}=0~~J^{ab}_{(1)}\eta_a \eta_b|_{\partial\mathcal{W}_{p+1}}=0~~,
\eeq
\beq\label{bound20}
\left[\nabla_{\hat{a}}J^{\hat{a}}_{(1)}-\eta_a\left(\hat{J}^{a}+p^{a\mu}{K^{a}}_{b\mu}\right)\right]|_{\partial\mathcal{W}_{p+1}}=0~~,
\eeq
where we have defined the boundary degrees of freedom $J^{\hat{a}}_{(1)}=J^{ab}_{(1)}v_{b}^{\hat{a}}\eta_a$ accounting for possible extra current sources on the brane boundary. Note that the structure $m^{\mu\nu}$ entering in the decomposition of ${J^{\mu\nu}_{(1)}}$ does not play a role in the equation of motion \eqref{motion20} neither in the boundary conditions \eqref{bound10}-\eqref{bound20} though it may be relevant when considering external couplings to background fields.

\subsubsection*{Extra symmetries and invariance of the equations of motion}
The current expansion \eqref{cpd} enjoys two symmetries as the stress-energy tensor \eqref{stpd} coined by the authors of \cite{Vasilic:2007wp} as `extra symmetry 1' and `extra symmetry 2'. Their transformation properties can be obtained by looking at the invariance of the functional
\beq \label{funct}
J[f]=\int d^{D}x\sqrt{-g}J^{\mu}f_{\mu}~~,
\eeq
for an arbitrary tensor field $f_{\mu}(x^{\alpha})$ of compact support. The `extra symmetry 1' is an exact symmetry to all orders in the expansion defined by the transformation
\beq\label{extra10}
\delta_1J^{\mu}_{(0)}=-\nabla_{a}\tilde\varepsilon^{\mu a}~~,~~\delta_1J^{\mu\nu}_{(1)}=-\tilde\varepsilon^{\mu a}u^{\nu}_{a}~~,
\eeq
and leaves the functional \eqref{funct} invariant as long as the parameters $\tilde\varepsilon^{\mu a}$ are required to obey $\tilde\varepsilon^{\mu a}\eta_a|_{\partial\mathcal{W}_{p+1}}=0$. This means, for example, that it is possible to gauge away one of the structures in the decomposition \eqref{decomp0} everywhere except on the boundary since $\delta_1(J^{\mu\nu}_{(1)}u^{a}_\nu)=-\tilde\varepsilon^{\mu a}$~. This is why we have left the last term in \eqref{decomp0} neither parallel nor orthogonal to the worldvolume. Further, invariance of the equation of motion \eqref{eom30} under \eqref{extra10} requires that $J^{ab}_{(1)}=J^{(ab)}_{(1)}$\footnote{In the case $q>0$ and also in the case of the stress-energy tensor \cite{Vasilic:2007wp}, the equations of motion obtained by this procedure are invariant under both `extra symmetries' without further requirements on the structures appearing in analogous decompositions to \eqref{decomp0}. The $q=0$ case is a special case as it is characterized by a current $J^{\mu}$ with only one index. This requires an extra constraint such that $J^{ab}_{(1)}=J^{(ab)}_{(1)}$ for the equations of motion \eqref{eom30} to be invariant under both extra symmetry transformations.}. Explicit use of \eqref{extra10} leads to the variations of the structures that characterize the charge current
\beq
\delta_1 \hat{J}^{a}=0~~,~~\delta_{1}p^{a\mu}=0~~,~~\delta_{1}J^{\hat{a}}_{(1)}=0~~,
\eeq
and hence leave the equations of motion \eqref{motion10}-\eqref{bound20} invariant. 

On the other hand, the `extra symmetry 2' is a perturbative symmetry and defined as the transformation that leaves \eqref{funct} invariant under the displacement of representative surface $X^{\alpha}(\sigma^{a})\to X^\alpha(\sigma^a)+\tilde\varepsilon^{\alpha}(\sigma^a)$. This leads to the transformation rule
\beq \label{extra20}
\delta_2 J^{\mu}_{(0)}=-J^{\mu}_{(0)}u^{a}_{\rho}\nabla_{a}\tilde\varepsilon^{\rho}-{\Gamma^{\mu}}_{\lambda\rho}J^{\lambda}_{(0)}\tilde\varepsilon^{\rho}~~,~~\delta_2 J^{\mu\rho}_{(1)}=-J^{\mu}_{(0)}\tilde\varepsilon^{\rho}~~.
\eeq
Explicit calculation using \eqref{extra20} leads to 
\beq \label{extra2_q_0}
\begin{split}
\delta_2 \hat{J}^{a}=-&J^{a}_{(0)}u^{b}_{\rho}\nabla_{b}\tilde\varepsilon^{\rho}-u^{a}_{\rho}J^{b}_{(0)}\nabla_{b}\tilde\varepsilon^{\rho}+\nabla_{b}\left(J^{a}_{(0)}\tilde\varepsilon^{b}\right)~~, \\
\delta_2p^{a\mu}&=-J^{a}_{(0)}\tilde\varepsilon^{\mu}~~,~~\delta_2J^{\hat{a}}_{(1)}=-J^{b}_{(0)}v_{b}^{\hat{a}}\tilde\varepsilon^{a}\eta_a~~,
\end{split}
\eeq
and renders the equations of motion \eqref{motion10}-\eqref{bound20} invariant. As we will be obtaining these tensor structures from specific black hole metrics in Sec.~\ref{sec:measurement}, the existence of this symmetry implies the existence of a residual gauge freedom in this measurement procedure as seen before in the case of the stress-energy tensor \cite{Armas:2012ac}.


\subsection{Equations of motion for branes carrying string charge} \label{sec:1charge}
In this section we give the equations of motion for branes carrying string charge ($q=1$) while a detailed derivation is given in App.~\ref{app:eom}. Such branes are characterized by the structures $J^{\mu\nu}_{(0)}$ and $J^{\mu\nu\rho}_{(1)}$. The same procedure as in the previous section allows us to split $J^{\mu\nu\rho}_{(0)}$ into the components
\beq\label{decomp1_1}
J^{\mu\nu\rho}_{(1)}=2u_{a}^{[\mu}m^{a\nu]\rho}+u_{a}^{\mu}u_{b}^{\nu}p^{ab\rho}+J^{\mu\nu a}_{(1)}u_{a}^{\rho}~~,
\eeq
where $m^{a\mu\nu}$ and $p^{ab\rho}$ are orthogonal in its space-time indices while $J^{\mu\nu a}_{(1)}$ is left neither parallel nor orthogonal to the worldvolume due to the extra symmetry transformations that we will describe bellow.  Similarly, for $J^{\mu\nu}_{(0)}$ we have the decomposition
\beq \label{decomp1_00}
J^{\mu\nu}_{(0)}=u^{\mu}_{a}u^{\nu}_{b}J^{ab}_{(0)}+2u_{b}^{[\mu}J^{\nu]b}_{\perp(1)}+J^{\mu\nu}_{\perp(1)}~~,
\eeq
for which the two last components are not independent but instead related to the dipole contributions in \eqref{decomp1_1} via the expressions
\beq \label{eom_1_q1}
J^{\mu b}_{\perp(1)}=u^{b}_{\rho}{\perp^{\mu}}_{\lambda}\nabla_{c}(J^{\rho\lambda c}_{(1)}-m^{c\rho\lambda})~~,~~J^{\mu\nu}_{\perp(1)}={\perp^{\mu}}_{\lambda}{\perp^{\nu}}_{\rho}\nabla_{c}(J^{\rho\lambda c}_{(1)}-m^{c\rho\lambda})~~.
\eeq
Using these relations we finally obtain the current conservation equation in the form
\beq\label{eom_2_q1}
\nabla_a\left(\hat{J}^{ab}-2p^{c[a(\mu)}{K_{c\mu}}^{b]}\right)=0~~,
\eeq
where the parenthesis in the index $\mu$ means that it is insensitive to the anti-symmetrization taking place only on the worldvolume indices. In Eq.~\eqref{eom_2_q1} we have introduced the effective worldvolume current $\hat{J}^{ab}=\hat{J}^{[ab]}$ such that
\beq
\hat{J}^{ab}=J^{ab}_{(0)}-u_{\mu}^{a}u_{\nu}^{b}\nabla_{c}J^{\mu\nu c}_{(1)}~~.
\eeq
Note again that, as in the case of branes carrying Maxwell charge, the components $m^{a\mu\nu}$ entering the decomposition \eqref{decomp1_1} do not play a role in the equation for current conservation. Similarly, for the boundary conditions we obtain
\beq \label{bound_1_q1}
\left(p^{ba\mu}+2{\perp^{\mu}}_{\lambda}J^{\lambda ab}_{(1)}\right)\eta_b|_{\partial\mathcal{W}_{p+1}}=0~~,~~J^{\mu ab}_{(1)}\eta_{a}\eta_b|_{\partial\mathcal{W}_{p+1}}=0~~,
\eeq
\beq \label{bound_2_q1}
\left[v^{b}_{\hat{b}}\nabla_{\hat{a}}J^{\hat{a}\hat{b}}_{(1)}-\eta_a\left(\hat{J}^{ab}-2p^{c[a\mu}{K_{c\mu}}^{b]}\right)\right]_{\partial\mathcal{W}_{p+1}}=0~~,
\eeq
where we have defined the boundary degrees of freedom $J^{\hat{a}\hat{b}}_{(1)}=J^{\mu \nu c}_{(1)}u_{c}^{\rho}\eta_{(\rho}v_{\nu)}^{\hat{a}}v_{\mu}^{\hat{b}}$~, with $v_{\mu}^{\hat{a}}=u_{\mu}^{a}v_{a}^{\hat{a}}$. Again, the components $m^{a\mu\nu}$ have dropped out of the boundary conditions. We see that branes carrying string charge are characterized by a worldvolume effective current $\hat{J}^{ab}$, a dipole moment $p^{ab\mu}$ and a boundary current $J^{\hat{a}\hat{b}}_{(1)}$.

\subsubsection*{Extra symmetries and invariance of the equations of motion}
The extra symmetries associated with the current \eqref{cpd} for branes carrying string charge are now deduced by looking at the functional
\beq \label{funct_q1}
J[f]=\int d^{D}x\sqrt{-g}J^{\mu\nu}f_{\mu\nu}~~,
\eeq
for an arbitrary tensor field $f_{\mu\nu}$ of compact support. The `extra symmetry 1' acts on the current \eqref{cpd} such that
\beq \label{extra1_q1}
\delta_1 J^{\mu\nu}_{(0)}=-\nabla_{a}\tilde\varepsilon^{\mu\nu a}~~,~~\delta_1 J^{\mu\nu\rho}_{(1)}=-\tilde\varepsilon^{\mu\nu a}u_{a}^{\rho}~~,
\eeq
where the parameters $\tilde\varepsilon^{\mu\nu a}$ satisfy the properties $\tilde\varepsilon^{\mu\nu a}=\tilde\varepsilon^{[\mu\nu] a}$ and $\tilde\varepsilon^{\mu\nu a}\eta_a|_{\partial\mathcal{W}_{p+1}}=0$~. This is turn implies that the components $J^{\mu\nu a}_{(1)}$ entering the decomposition \eqref{decomp1_1} are pure gauge everywhere except on the boundary since $\delta_1(J^{\mu\nu\rho}_{(1)}u_{\rho}^{a})=J^{\mu\nu a}_{(1)}=-\tilde\varepsilon^{\mu\nu a}$. Evaluating \eqref{extra1_q1} for the components that describe the charged pole dipole brane leads to the variations
\beq
\delta_{1}\hat{J}^{ab}_{(0)}=0~~,~~\delta_1 p^{ab\rho}=0~~,~~\delta_1 J^{\hat{a}\hat{b}}_{(1)}=0~~,
\eeq
and hence the equations of motion \eqref{eom_1_q1}-\eqref{eom_2_q1} together with the boundary conditions \eqref{bound_1_q1}-\eqref{bound_2_q1} are invariant under this transformation rule. Turning our attention to the 'extra symmetry 2', invariance of \eqref{funct_q1} requires
\beq \label{extra2_q1}
\delta_{2}J^{\mu\nu}_{(0)}=-J^{\mu\nu}_{(0)}u^{a}_{\rho}\nabla_{a}\tilde\varepsilon^{\rho}-2{\Gamma^{[\mu}}_{\rho\lambda}J^{\nu]\lambda}_{(0)}\thinspace\tilde\varepsilon^{\rho}~~,~~\delta_2J^{\mu\nu\rho}_{(1)}=-J^{\mu\nu}_{(0)}\tilde\varepsilon^\rho~~,
\eeq
which upon explicit calculation leads to
\beq
\begin{split}
\delta_2 \hat{J}^{ab}=-&J^{ab}_{(0)}u^{c}_{\rho}\nabla_{c}\tilde\varepsilon^{\rho}-2u^{[a}_{\rho}J^{b] c}_{(0)}\nabla_{c}\tilde\varepsilon^{\rho}+\nabla_{c}\left(J^{ab}_{(0)}\tilde\varepsilon^{c}\right)~~, \\
\delta_2p^{ab\mu}&=-J^{ab}_{(0)}\tilde\varepsilon^{\mu}~~,~~\delta_2J^{\hat{a}\hat{b}}_{(1)}=-J^{ab}_{(0)}\tilde\varepsilon^{c}u_{c}^{\rho}\eta_{(\rho}v_{\nu)}^{\hat{a}}v_{\mu}^{\hat{b}}~~,
\end{split}
\eeq
and leaves the equations of motion \eqref{eom_1_q1}-\eqref{eom_2_q1} and boundary conditions \eqref{bound_1_q1}-\eqref{bound_2_q1} invariant.


\subsection{Equations of motion for branes charged under higher-form fields}
In this section we conjecture the equations of motion for the cases $q>1$ and leave the proof for later work. In these cases, the electric current of the $p$-brane are characterized by two structures $J^{\mu_1...\mu_{q+1}}_{(0)}$ and $J^{\mu_1...\mu_{q+1}\rho}_{(1)}$. It is straight forward to derive a similar constraint as in Eq.~\eqref{eom10} by solving the conservation equation \eqref{ceom}. This constraint allows us to make the general decomposition of $J^{\mu_1...\mu_{q+1}\rho}_{(1)}$ such that
\beq \label{decompq_q}
J^{\mu_{1}...\mu_{q+1}\rho}_{(1)}=[(q+1)!/q!]u_{a_1}^{[\mu_1}...u_{a_q}^{\mu_{q}}m^{(a_1)...(a_q)\mu_{q+1}]\rho}+u_{a_1}^{\mu}...u_{a_{q}}^{\mu_{q}}u^{\mu_{q+1}}_{b}p^{a_{1}...a_{q+1}\rho}+J^{\mu_{1}...\mu_{q+1} a}_{(1)}u_{a}^{\rho}~~,
\eeq
where the parenthesis on each of the indices $a_1...a_q$ indicates that these indices are insensitive to the anti-symmetrization which is done only over the space-time indices $\mu_{1}...\mu_{q+1}$. Moreover, $m^{a_1...a_q\mu_{q+1}\rho}$ satisfies the properties $m^{a_1...a_q\mu_{q+1}\rho}=m^{[a_1...a_q]\mu_{q+1}\rho}=m^{a_1...a_q[\mu_{q+1}\rho]}$ while $p^{a_{1}...a_{q+1}\rho}$ has the property $p^{a_{1}...a_{q+1}\rho}=p^{[a_{1}...a_{q+1}]\rho}$ and also $J^{\mu_{1}...\mu_{q+1} a}_{(1)}=J^{[\mu_{1}...\mu_{q+1}] a}_{(1)}$. A similar decomposition of $J^{\mu_1...\mu_{q+1}}_{(0)}$ as in \eqref{decomp1_00} is assumed and the final form of the equations of motion is conjectured to be
\beq\label{eom_2_qq}
\nabla_{a_1}\left(\hat{J}^{a_1...a_{q+1} }+(-1)^{q} [(q+1)!/q!] \thinspace p^{c[a_1...a_q(\mu) }{K_{c\mu}}^{a_{q+1}] }\right)=0~~,
\eeq
while the boundary conditions have an analogous form to the $q=1$ case presented in the previous section
\beq \label{bound_1_qq}
\left(p^{a_1...a_{q+1}\mu}+(-1)^{q}q!{\perp^{\mu}}_{\lambda}J^{\lambda a_{1}...a_{q+1}}\right)\eta_{a_{q+1}}|_{\partial\mathcal{W}_{p+1}}=0~~,~~J^{\mu_{1}...\mu_{q} a_{q+1}b}_{(1)}\eta_{a_{q+1}}\eta_b|_{\partial\mathcal{W}_{p+1}}=0~~,
\eeq
\beq \label{bound_2_qq}
\left[v^{b}_{\hat{b}}\nabla_{\hat{a}_1}J^{\hat{a}_1...\hat{a}_{q+1}}_{(1)}-\eta_{a_1}\left(\hat{J}^{a_1...a_{q+1}}+(-1)^q [(q+1)!/q!]\thinspace p^{c[a_1...a_q(\mu) }{K_{c\mu}}^{a_{q+1}] }\right)\right]_{\partial\mathcal{W}_{p+1}}=0~~.
\eeq
Here we have introduced the effective worldvolume current $\hat{J}^{a_1...a_{q+1}}=J^{a_{1}...a_{q+1}}_{(0)}-u_{\mu}^{a_1}...u_{\mu_{q+1}}^{a_{q+1}}\nabla_{c}J^{\mu_1...\mu_{q+1} c}_{(1)}$, as well as the boundary degrees of freedom $J^{\hat{a}_1...\hat{a}_{q+1}}_{(1)}=J^{a_1...a_{q-1}\mu_q\mu_{q+1}c}_{(1)}u_{c}^{\rho}\eta_{(\rho}v_{\mu_{q+1})}^{\hat{a}_{q}}v_{\mu_{q}}^{\hat{a}_{q+1}}$. The conjectured form of the equations of motion \eqref{eom_2_qq}-\eqref{bound_2_qq} for any value of $q$ is supported by their invariance under the extra symmetry transformations which we will now analyze. 

The `extra symmetry 1' acting on the generic form of the current \eqref{cpd} has the following transformation rule
\beq \label{extra1_qq}
\delta_1 J^{\mu_1...\mu_{q+1}}_{(0)}=-\nabla_{a}\tilde\varepsilon^{\mu_{1}...\mu_{q+1} a}~~,~~\delta_1 J^{\mu_1...\mu_{q+1}\rho}_{(1)}=-\tilde\varepsilon^{\mu_{1}...\mu_{q+1} a}u_{a}^{\rho}~~,
\eeq
where $\tilde\varepsilon^{\mu_{1}...\mu_{q+1} a}$ has the property $\tilde\varepsilon^{\mu_{1}...\mu_{q+1} a}=\tilde\varepsilon^{[\mu_{1}...\mu_{q+1}] a}$ and is constrained on the boundary such that $\tilde\varepsilon^{\mu_{1}...\mu_{q+1} a}\eta_a|_{\partial\mathcal{W}_{p+1}}=0$. Correspondingly, this implies that the structures characterizing the charged pole-dipole brane transform as $\delta_1\hat{J}^{a_1...a_{q+1}}=\delta_{1}p^{a_1...a_{q+1}\mu}=\delta_{1}J^{\hat{a}_{1}...\hat{a}_{q+1}}=0$, leaving the equations of motion invariant. As in the cases $q=0$ and $q=1$ analyzed previously, this symmetry implies that the last structure introduced in \eqref{extra1_qq} can be gauged away everywhere on the worldvolume. Furthermore, under the action of the `extra symmetry 2' the structures entering \eqref{cpd} transform according to
\beq
\delta_{2}J^{\mu_1...\mu_{q+1}}_{(0)}=-J^{\mu_1...\mu_{q+1}}_{(0)}u^{c}_{\rho}\nabla_{c}\tilde\varepsilon^{\rho}-[(q+1)!/q!]{\Gamma^{[\mu_1}}_{\rho\lambda}J^{\mu_2...\mu_{q+1}]\lambda}_{(0)}\thinspace\tilde\varepsilon^{\rho}~~,~~\delta_2J^{\mu_1...\mu_{q+1}\rho}_{(1)}=-J^{\mu_1...\mu_{q+1}}_{(0)}\tilde\varepsilon^\rho~~,
\eeq
which upon explicit calculation lead to the `extra symmetry 2' transformations
\beq
\begin{split}
\delta_2 \hat{J}^{a_1...a_{q+1}}=-&J^{a_{1}...a_{q+1}}_{(0)}u^{c}_{\rho}\nabla_{c}\tilde\varepsilon^{\rho}-[(q+1)!/q!]u^{[a_1}_{\rho}J^{a_2...a_{q+1}] c}_{(0)}\nabla_{c}\tilde\varepsilon^{\rho}+\nabla_{c}\left(J^{a_1...a_{q+1}}_{(0)}\tilde\varepsilon^{c}\right)~~, \\
\delta_2p^{a_1...a_{q+1}\mu}&=-J^{a_1...a_{q+1}}_{(0)}\tilde\varepsilon^{\mu}~~,~~\delta_2J^{\hat{a}_1...\hat{a}_{q+1}}_{(1)}=-J^{a_1...a_{q+1}}_{(0)}\tilde\varepsilon^{c}u_{c}^{\rho}\eta_{(\rho}v_{\nu)}^{\hat{a}}v_{\mu}^{\hat{b}}~~.
\end{split}
\eeq
The above transformation rules leave the equations of motion \eqref{eom_2_qq}-\eqref{bound_2_qq} invariant. In Sec.~\ref{sec:measurement} we will give examples of these structures measured for bent black branes.


\section{Physical interpretation and brane electroelasticity \label{sec:physical} }

In this section we discuss the physical interpretation of the structures entering the dipole contribution of the stress-energy
tensor and the electric current. 
The physical interpretation of the structures $j^{b\mu\nu}$ and $d^{ab\rho}$ introduced in Eq.~\eqref{decomp_T} 
was given already in Refs.~\cite{Armas:2011uf, Armas:2013hsa} and will be reviewed here. 
Furthermore, the physical interpretation of the different structures appearing in the decomposition of the electric  current \eqref{decomp0} for the $q=0$ case was in part discussed already in \cite{Armas:2012ac}, while here we also present
the generalization relevant to the cases \eqref{decomp1_1} ($q=1$) and \eqref{decompq_q} ($q>1$) studied in the previous section. 

\subsection{Bending moment and Young modulus}

As mentioned in the beginning of Sec.~\ref{sec:dynamics} the structure $d^{ab\rho}$ accounts for the bending moment of the 
brane \cite{Armas:2011uf, Armas:2013hsa}.  To see this note that we can compute the total bending moment from the stress-energy tensor 
\eqref{stpd} by 
\beq
\label{bendmom}
D^{ab \rho }=\int_{\Sigma}d^{D-1}x\sqrt{-g} T^{\mu \nu }u_\mu^a u_\nu^b x^{\rho } =
\int_{\mathcal{B}_p}d^{p}\sigma  \sqrt{-\gamma} d^{a b \rho } ~~, 
\eeq
where $\Sigma$ is a constant timeslice in the bulk space time and we have ignored boundary terms (which we will continue
to do so in the following). Hence we identify $d^{a b \rho} $ as the bending moment density on the brane. 
Note that for the case of a point particle ($p=0$), $d^{ab\rho}$ has only one non-vanishing component, namely, $d^{\tau\tau\rho}$ where $\tau$ is the proper time coordinate of the worldline. Since the stress-energy tensor \eqref{stpd} also enjoys the `extra symmetry 2' acting as $\delta_2 d^{ab\rho}=-T^{ab}_{(0)}\tilde\varepsilon^{\rho}$ one can, by an appropriate choice of $\tilde\varepsilon^\rho$, gauge away the component $d^{\tau\tau\rho}$ \cite{Armas:2011uf}. Thus point particles do not carry worldvolume mass dipoles but in the case $p>0$ these components cannot, in general, be gauged away.

The bending moment $d^{ab\rho}$ is a priori unconstrained but assuming that the brane will behave according to classical (Hookean) elasticity theory we consider it to be of the form
\beq \label{eqn:YMdef}
d^{ab\rho}=\tilde Y^{abcd}{K_{cd}}^{\rho}~~,
\eeq
which is the form of the bending moment expected for a thin elastic brane that has been subject to pure bending. Here, the extrinsic curvature ${K_{cd}}^{\rho}$ has the interpretation of the Lagrangian strain since it measures the variation of the induced metric on the brane along transverse directions to the worldvolume while $\tilde Y^{abcd}$ is the Young modulus of brane
\footnote{We use the convention that  $\tilde Y = Y I $ (omitting tensor indices) where $Y$ is the conventionally normalized Young modulus and $I$ the moment inertia of the object with respect to the choice of worldvolume surface.}.
 
The application of this paper is to bending deformations of fluid branes which are stationary. In these situations the general structure of $\tilde Y^{abcd}$ has been classified for neutral isotropic fluids using an effective action approach \cite{Armas:2013hsa}. For the isotropic cases studied here, making a slight generalization to the case of $p$-branes with
worldvolume 0-brane (Maxwell) charge, it takes the form \cite{Armas:2013hsa}\footnote{Note that the Young modulus $\tilde Y^{abcd}$ introduced here is related to the one introduced in \cite{Armas:2013hsa} via the relation $\tilde Y^{abcd}=-\mathcal{Y}^{abcd}$.}
\beq \label{YM_0}
\begin{split}
\tilde Y^{abcd}= 
-2 & \left(  \lambda_{1}(\textbf{k}; T,\Phi_H) \gamma^{ab} \gamma^{cd} + 
\lambda_{2}(\textbf{k};T,\Phi_{H})\gamma^{a(c}\gamma^{d)b}
+\lambda_{3}(\textbf{k};T,\Phi_{H})\textbf{k}^{(a}\gamma^{b)(c}\textbf{k}^{d)} \right. \\ 
& \left. +\lambda_{4}(\textbf{k};T, \Phi_{H}) \frac{1}{2} ( \textbf{k}^{a}\textbf{k}^{b} \gamma^{cd} + \gamma^{ab}
\textbf{k}^{c}\textbf{k}^{d} ) 
+\lambda_{5}(\textbf{k};T, \Phi_{H})\textbf{k}^{a}\textbf{k}^{b} \textbf{k}^{c}\textbf{k}^{d} \right)~,
\end{split}
\eeq
where $\textbf{k}^{a}$ is the Killing vector field along which the fluid is moving with  $\textbf{k}=|-\gamma_{ab}\textbf{k}^{a}\textbf{k}^{b}|^{\frac{1}{2}}$. We have also indicated explicitly the dependence on the global temperature $T$, 
and the generalization compared to the neutral isotropic case of \cite{Armas:2013hsa} is that there
is  now in addition  a dependence on the global chemical potential $\Phi_H$. 
The Young modulus $\tilde Y^{abcd}$ satisfies the expected properties of a classical elasticity tensor $\tilde Y^{abcd}=\tilde Y^{(ab)(cd)}=\tilde Y^{cdab}$. 
We will find explicit realizations of \eqref{YM_0} in Sec.~\ref{sec:branesq0} when we consider the bending of black $p$-branes with 0-brane
charge. For this we note that not all of the five terms in the expression in \eqref{YM_0} are independent. In fact, due to the `extra symmetry 2', 
these include gauge dependent terms \cite{Armas:2013hsa} of the form  $k\left(T^{ab}_{(0)}\gamma^{cd}+T^{cd}_{(0)}\gamma^{ab}\right)$, so that in the end only three out of the five $\lambda$-coefficients are independent, when using also the equations of motion.

It is not the purpose of this work to construct the effective action for anisotropic fluid branes. However, we note that 
the simplest case of $p$-branes carrying $1$-brane (string) charge with $p>1$,  are characterized by a vector $v^{a}$ satisfying $v^{a}u_{a}=0$ and $v^{a}v_{a}=1$, aligned in the direction of the smeared $1$-brane charge along the brane worldvolume. Following the analysis of \cite{Armas:2013hsa}, there are four further response coefficients that can in principle be added to the effective action and, in turn, to the Young modulus defined in \eqref{YM_0}, restricting to terms that contain only
even powers of $u^a$ and/or $v^a$. The expression in \eqref{YM_0} should therefore
be  supplemented with a contribution of the form
\beq \label{YM_1}
\begin{split}
\hat Y^{abcd}=-2 & \left(
\lambda_{6}(\textbf{k}, \bv; T,\Phi_H) \textbf{k}^{a}\textbf{k}^{b}\bv^{c}\bv^{d}
+\lambda_{7}(\textbf{k}, \bv; T,\Phi_H)\bv^{a}\bv^{b}\textbf{k}^{c}\textbf{k}^{d} \right. \\
&
\left. 
+\lambda_{8}(\textbf{k},\bv ; T,\Phi_H)\textbf{k}^{(a}\bv^{b)}\textbf{k}^{(c}\bv^{d)}
+\lambda_{9}(\textbf{k},\bv; T,\Phi_H)\bv^{a}\bv^{b}\bv^{c}\bv^{d}  
\right)
~~.
\end{split}
\eeq
Here we have introduced the non-normalized space-like vector $\bv^a$ in terms of which $v^a = \bv^a/ \bv $ where
$\bv = | \bv^a \bv^b \gamma_{ab} |^{1/2}$. Note that for this case the $\lambda$-coefficients introduced in \eqref{YM_0} are now also functions of $\bv$.
More generally, for branes with smeared $q$-brane charge with $q>0$ and $p>q$ one can introduce a set of vectors 
$v^{a}_{(i)}~,~i=1,...,q$~, such that $v^{a}_{(i)}v_a^{(j)}=\delta_i^j$ and furthermore $v^{a}_{(i)}u_{a}=0$. 
As a consequence, one can have for every vector $v^{a}_{(i)}$, a contribution of the form \eqref{YM_1}, but clearly
more complicated contributions can appear as well. The analysis of this is beyond the scope of the present paper. 

We note that the introduction of the new terms in \eqref{YM_1} does not a priori guarantee that the expected classical symmetries $\hat Y^{abcd}=\hat Y^{(ab)(cd)}=\hat Y^{cdab}$ are preserved.  However, one should properly take into account that just as in \eqref{YM_0}, as a consequence of gauge freedom not all of the terms in \eqref{YM_1} are independent. In fact, we will see that for the particular cases
of  charged black branes with $q>0$ considered in this paper, the terms appearing in \eqref{YM_1} can be transformed
away, so that we will find that in all our cases the Young modulus is described by the expression \eqref{YM_0}. 
In Sec.~\ref{sec:measurement}, we will give explicit examples of charged black branes  exhibiting these properties.
It would be interesting to examine whether more general bent charged black brane solutions can be constructed that
necessitate the introduction of the terms in \eqref{YM_1}, and, moreover,  whether in those cases there is
an anomalous contribution violating the classical symmetries mentioned above.

\subsection{Electric dipole moment and piezoelectric moduli}

We now proceed to interpret the structures entering in the decomposition of the electric current for the different cases,
focussing first on the quantity $p^{a  \rho}$ relevant to the case  $q=0$ for which we have the current defined
in \eqref{cpd}.  In close parallel to the bending moment in \eqref{bendmom},  a charged brane 
 can have an electric dipole moment  $P^{a\rho}$ due to the finite thickness.
This is obtained by evaluating 
\beq
\label{eldipole}
P^{a\rho}=\int_{\Sigma}d^{D-1}x\sqrt{-g}\thinspace J^{\mu}u_{\mu}^{a}x^{\rho}=\int_{\mathcal{B}_p}d^{p}\sigma\sqrt{-\gamma}p^{a\rho}~~,
\eeq
and hence the structure $p^{a\rho}$ should be interpreted as a density of worldvolume electric dipole moment.
Note that in the case of a point particle the structure $p^{a\rho}$ appearing in the decomposition \eqref{decomp0} can be gauged away due to the `extra symmetry 2' since it only has one worldvolume index component $p^{\tau\rho}$ where $\tau$ is the proper time direction of the worldline. Since by Eq.~\eqref{extra2_q_0} we have that $\delta_2 p^{\tau\rho}=-J^{\tau}_{(0)}\tilde\varepsilon^{\rho}$~, one can always choose $\tilde\varepsilon^{\rho}$ such that the component $p^{\tau\rho}$ vanishes. Therefore, point particles cannot carry worldvolume electric dipoles in the same way that they cannot carry worldvolume mass dipoles. 
For extended objects ($p\ge1$) the electric dipole moment cannot be removed generically so the brane will have an electric dipole moment . 

We now specialize to a class of branes for which the form of $p^{a\rho}$ is that expected from classical electroelasticity theory
\beq
\label{eq:piezo}
p^{a\rho}=\tilde \kappa^{abc}{K_{bc}}^{\rho}~~,
\eeq
which is the covariant generalization of the usual relation for the electric dipole moment of 
classical piezoelectrics \cite{bardzokas2007mathematical}. Here, $\tilde \kappa^{abc}$ is a set of piezoelectric moduli encoding the brane response to bending deformations. The structure of $\tilde \kappa^{abc}$ has not been yet classified from an effective action perspective as was
the case for  \eqref{YM_0}. However, based on covariance, it is easy to write down the expected form for the 
cases we consider such that $\tilde \kappa^{abc}$ obeys the symmetry property $\tilde \kappa^{abc}=\tilde \kappa^{a(bc)}$ and respects the gauge freedom set by the transformation rule \eqref{extra2_q_0}. This leads to the form
\beq \label{PM_0}
\tilde \kappa^{abc}=-2\left(\kappa_1(\textbf{k};T,\Phi_H)\gamma^{a(b}\textbf{k}^{c)}+\kappa_2(\textbf{k};T,\Phi_H)\textbf{k}^{a}\textbf{k}^{b}\textbf{k}^{c} + \kappa_3(\textbf{k};T,\Phi_H) \textbf{k}^{a} \gamma^{bc} \right)~~.
\eeq
In parallel with \eqref{YM_0} this contains gauge-dependent terms with respect to the `extra symmetry 2', which have the form
$k \thinspace J^{a}_{(0)}\gamma^{bc}$, with $k$ the gauge parameter. In all, there is only one independent $\kappa$-parameter
 when using also the equations of motion. 

We now turn our attention to the case of general $p$-branes carrying smeared $q$-brane charge with $q>0$. 
The generalization of  \eqref{eldipole} is the  electric dipole moment $P^{a_1...a_{q+1}\rho}$ defined by 
\beq
P^{a_1...a_{q+1}\rho}=\int_{\Sigma}d^{D-1}x\sqrt{-g}\thinspace J^{\mu_{1}...\mu_{q+1}\rho}u_{\mu_1}^{a_1}...u^{a_{q+1}}_{\mu_{q+1}}x^{\rho}=\int_{\mathcal{B}_p}d^{p}\sigma\sqrt{-\gamma}p^{a_1...a_{q+1}\rho}~~,
\eeq
and hence $p^{a_1...a_{q+1}\rho}$ has the same interpretation as for the $q=0$ case. Now according to the expectation from classical electro-elastodynamics we assume the following form for $p^{a_1...a_{q+1}\rho}$
\beq
\label{eq:piezo2}
p^{a_1...a_{q+1}\rho}=\tilde \kappa^{a_1...a_{q+1}bc}{K_{bc}}^{\rho}~~,
\eeq
where $\tilde \kappa^{a_1...a_{q+1}bc}$ inherits the symmetries of $p^{a_1...a_{q+1}\rho}$, that is, $\tilde \kappa^{a_1...a_{q+1}bc}=\tilde \kappa^{[a_1...a_{q+1}]bc}$ and also the property $\tilde \kappa^{a_1...a_{q+1}bc}=\tilde \kappa^{a_1...a_{q+1}(bc)}$. In particular, for $q=1$, one expects a structure of the form
\beq \label{PM_1}
\begin{split}
\tilde \kappa^{abcd}=-2 & \left(\kappa_1(\textbf{k},\bv;T,\Phi_H)\bv^{[a}\gamma^{b](c}\textbf{k}^{d)}+\kappa_2(\textbf{k},\bv ; T,\Phi_H)\bv^{[a}\textbf{k}^{b]}\textbf{k}^{c}\textbf{k}^{d} \right. \\
 & \left. +\kappa_3(\textbf{k},\bv; T, \Phi_H)\bv^{[a}\textbf{k}^{b]}\bv^{c}\bv^{d} 
+  \kappa_4(\textbf{k},\bv; T, \Phi_H) \bv^{[a} \textbf{k}^{b]} \gamma^{cd} \right)~~. 
\end{split}
\eeq
Again, this includes a gauge-dependent term of the form $k \thinspace J^{ab}_{(0)}\gamma^{cd}$. 
The symmetry property of the piezoelectric moduli $\tilde \kappa^{abcd}$ for $q\ge1$, namely the anti-symmetry in its first $q+1$ indices is not something that has a classical analogue and has not been previously considered in the literature of charged elastic solids. In Sec.~\ref{sec:measurement} we will give examples of $\tilde \kappa^{a_1...a_{q+1}bc}$ measured from charged black branes in gravity. In particular, we will find that for all the cases considered in this paper, there is only
one independent contribution.  

\subsection{Spin current and magnetic dipole moment \label{sec:spincurrent} }

As mentioned in the beginning of Sec.~\ref{sec:dynamics} the structure  $j^{b\mu\nu}$ accounts for the spinning degrees of freedom of the brane \cite{Armas:2011uf, Armas:2013hsa}.  This can be seen by using the stress-energy tensor in 
\eqref{stpd} and constructing the total angular momentum in a
$(\mu,\nu)$-plane orthogonal to the brane as 
\beq
J_\perp^{\mu\nu}=\int_{\Sigma}d^{D-1}x\sqrt{-g}\left(T^{\mu 0 }x^{\nu}-T^{\nu 0 }x^{\mu}\right)=
\int_{\mathcal{B}_p}d^{p}\sigma  \sqrt{-\gamma} j^{0 \mu \nu} ~~.
\eeq
 Hence we recognize $j^{b \mu \nu}$ as the angular momentum density on the brane. Angular momentum conservation follows
 because the brane worldvolume spin current $j^{b \mu \nu}$ is conserved \cite{Armas:2011uf, Armas:2013hsa}. As we will now see, this quantity
is also expected to play a role in relation to a particular component of the dipole contribution to the electric current 
 for branes with  $q$-charge.
  
For this we first turn to the quantity $m^{\mu \nu}$ entering the decomposition of the electric current for $q=0$.  Here, it 
is instructive to furthermore start by considering the case of a point particle ($p=0$) with point-like charge. This can have a magnetic dipole moment $M^{\mu\nu}$ obtained by evaluating
\beq
M^{\mu\nu}=\int_{\Sigma}d^{D-1}x\sqrt{-g}\left(J^{\mu}x^{\nu}- J^{\nu}x^{\mu}\right)=\int_{\mathcal{B}_p}d^{p}\sigma\sqrt{-\gamma}m^{\mu\nu}~~.
\eeq
Therefore, $m^{\mu\nu}$ should be seen as a worldvolume density of magnetic dipole moment. Since a magnetic dipole moment requires a moving charge, the most natural thing is to expect $m^{\mu\nu}$ to be proportional to the spin current
 $j^{\tau\mu\nu}$ of the particle. This interpretation also holds for any $p$-brane with smeared $0$-brane charge and generically one should expect
\beq
\label{magmom}
m^{\mu\nu}=\lambda(\sigma^{b}) u_{a}j^{a\mu\nu}~~,
\eeq
for some worldvolume function $\lambda(\sigma^{b})$. Turning to the case of general $p$-branes carrying a smeared $q$-brane charge with $q>0$, we can evaluate the magnetic dipole moment
\beq
M^{a_1...a_q\mu\nu}=\int_{\Sigma}d^{D-1}x\sqrt{-g}\left(J^{\mu_1...\mu_q\mu}x^{\nu}-J^{\mu_1...\mu_q\nu}x^{\mu}\right)u_{\mu_1}^{a_1}...u_{\mu_{q}}^{a_q}=\int_{\mathcal{B}_p}d^{p}\sigma\sqrt{-\gamma}m^{a_1...a_q\mu\nu}~~,
\eeq
and hence generically the structure $m^{a_1...a_q\mu\nu}$ should be interpreted as a density of magnetic dipole moment.
Moreover, in analogy with \eqref{magmom} we expect this to be related to the spin current via the generic form 
\beq
m^{a_1...a_q\mu\nu}={\Xi^{a_1...a_q}}_{b}j^{b\mu\nu}~~,
\eeq
where ${\Xi^{a_1...a_q}}_{b}$ is totally anti-symmetric in its indices $a_1...a_q$. We will not find explicit examples
of these responses to the spin in this paper, since the  black branes that we consider in Sec.~\ref{sec:measurement}
are non-spinning.


\section{Measuring the response coefficients} \label{sec:measurement}

In this section we construct a class of bent charged black brane geometries and measure their response coefficients.
These provide explicit realizations in (super)gravity theories of the general results for charged fluid branes presented in Secs.~\ref{sec:dynamics} and  \ref{sec:physical}. We begin by describing the framework for obtaining the response coefficients characterized in Sec.~\ref{sec:physical} from the large $r$-asymptotics of a charged black brane solution. We then describe in detail the solution generating techniques used in order to obtain large classes of charged bent metrics. Finally, we provide the thermodynamics, the Young modulus \eqref{YM_0} and the piezoelectric moduli \eqref{PM_0}, \eqref{PM_1} of the solutions constructed using these techniques.

\subsection{Setup and large \texorpdfstring{$r$}{r}-asymptotics}

We consider asymptotically flat charged dilatonic black brane solutions of the theory with action
\begin{equation} \label{eqn:action}
	S = \frac{1}{16\pi G} \int d^D x \sqrt{-g} \left[ R - \frac{1}{2} (\nabla \phi)^2 - \frac{1}{2(q+2)!} e^{a\phi} H_{[q+2]}^2 \right] ~~,
\end{equation}
where we note that the field content consists of the metric $g_{\mu\nu}$, the $(q+1)$-form gauge potential $A_{[q+1]}$ with 
field strength $H_{[q+2]} = \text{d}A_{[q+1]}$ and the dilaton $\phi$. The measurement of the Young modulus and the piezoelectric moduli of bent black branes has been considered previously in the literature. The Young modulus of neutral black strings was first measured in \cite{Armas:2011uf} and later extended to black $p$-branes in \cite{Camps:2012hw}. In App.~\ref{app:neutral} we review these results in detail and provide further details on the notation used here. The first example of the piezoeletric moduli was measured for the charged black string in EMD theory \cite{Armas:2012ac} and here we will extend this analysis to large classes of bent charged $p$-branes that
are solutions of the action \eqref{eqn:action}. 
In particular, we will consider a subset of solutions of the action \eqref{eqn:action} which falls into the class of the generalized Gibbons-Maeda black brane family of solutions found in Ref.~\cite{Caldarelli:2010xz}, describing $p$-branes with smeared $q$-brane charge
with $p\geq q$. In Sec.~\ref{sec:solutions}, we will provide the details on how the bent versions of these solutions are constructed.
 
We now outline the method used to measure these response coefficients for generic black brane solutions in the theory described by \eqref{eqn:action}. As mentioned in the beginning of Sec.~\ref{sec:dynamics}, bent branes acquire a bending moment which in turn implies a dipole correction $T^{\mu\nu\rho}_{(1)}$ to the stress-energy tensor and if the brane is charged an electric dipole moment $J^{\mu_{1}...\mu_{q+1}\rho}_{(1)}$ is also induced. To measure these from a gravitational solution in a theory with action \eqref{eqn:action} we look at its large $r$-asymptotics where the geometry and gauge field, as seen from a distant observer, can be replaced by effective sources of stress-energy and current. The task is then to find the effective stress-energy tensor \eqref{stpd} and current \eqref{cpd} that source the charged brane solution. To this end, we note that the equations of motion that follow from the action \eqref{eqn:action} in the presence of sources are given by
\begin{equation} \label{eqn:einsteineqn}
	G_{\mu\nu} - \frac{1}{2} \nabla_{\mu} \phi \nabla_{\nu} \phi 
	- \frac{1}{2(q+1)!} e^{a\phi} \left(  H_{\mu \rho_1 \ldots \rho_{q+1}} H_{\nu}^{\phantom{\mu}\rho_1 \ldots \rho_{q+1}} - \frac{1}{2(q+2)}  H^2 g_{\mu\nu} \right) = 8 \pi G T_{\mu\nu} \,,
\end{equation}
\begin{equation} \label{eqn:gaugefieldeom}
	\nabla_{\nu} \left( e^{a\phi} H^{\nu \mu_1 \ldots \mu_{q+1}} \right) = -16\pi G J^{\mu_1 \ldots \mu_{q+1}} \,, \qquad
	\Box \phi - \frac{a}{2(q+2)!} e^{a\phi} H^2 = 0 ~~,
\end{equation}
where $T^{\mu\nu}$ and $J^{\mu_1...\mu_{q+1}}$ are the effective stress-energy tensor and current given in Eq.~\eqref{stpd} and Eq.~\eqref{cpd} respectively and encode the brane dynamics far from the black brane horizon. The bending moment \eqref{eqn:YMdef} and the electric dipole moment \eqref{eq:piezo2} are then related, via Eqs.~\eqref{eqn:einsteineqn}-\eqref{eqn:gaugefieldeom}, to the dipole corrections occurring in the different fields as one approaches spatial infinity, which by definition have the fall-off behaviour $\mathcal{O}\left(r^{-n-1}\right)$ \cite{Armas:2011uf}. In particular, the bending moment is related to the dipole contributions to the metric $g_{\mu\nu}$ far away from the brane horizon. It is therefore convenient to decompose the metric according to
\begin{equation} \label{eqn:decomp}
	g_{\mu\nu} = \eta_{\mu\nu} + h^{(M)}_{\mu\nu} + h^{(D)}_{\mu\nu} + \mathcal{O}(r^{-n-2}) ~~,
\end{equation}
where the coefficients $h^{(M)}_{\mu\nu}$ represent the monopole structure of the metric, generically of order $\mathcal{O}\left(r^{-n}\right)$, while the coefficients $h^{(D)}_{\mu\nu}$ represent the dipole deformation of the metric of order $\mathcal{O}\left(r^{-n-1}\right)$. Similarly, the electric dipole moment is related to the dipole contributions to the gauge field $A_{\mu_{1}...\mu_{q+1}}$, also of order $\mathcal{O}\left(r^{-n-1}\right)$. Therefore we decompose the gauge field such that
\begin{equation} \label{eqn:decompB}
	A_{\mu_{1}...\mu_{q+1}} = A_{\mu_{1}...\mu_{q+1}}^{(M)} + A_{\mu_{1}...\mu_{q+1}}^{(D)} + \mathcal{O}\left( r^{-n-2} \right) ~~,
\end{equation}
where again the labels $(M)$ and $(D)$ indicate the monopole and dipole contributions respectively to the gauge field $A_{\mu_{1}...\mu_{q+1}}$. We note that in the cases studied here, there are no response coefficients associated with the dilaton $\phi$, a fact that renders the analysis of the dilaton equation of motion given in \eqref{eqn:einsteineqn} unnecessary.
In the following, we will review how the bending and electric dipole moments as well as the corresponding response coefficients  can be extracted from the linearized equations of motion. We should emphasize that the procedure that will be outlined here only works under the assumption that there are no background fields, namely, no background gauge field nor a non-zero background dilaton and that the background metric is asymptotically flat.

\subsubsection*{Measuring the Young modulus} \label{sec:ComputingYM}

Expanding the r.h.s. of Eq.\eqref{eqn:einsteineqn} according to Eq.~\eqref{stpd} and using the decomposition \eqref{decomp_T}, one finds that the dipole contribution to the metric should satisfy the linearized equation of motion
\begin{equation} \label{eqn:linEin}
	\nabla^2_{\perp} \bar{h}^{(D)}_{\mu\nu} = 16\pi G d^{\phantom{\mu\nu}r_{\perp}}_{\mu\nu} \partial_{r_{\perp}}  \delta^{n+2}(r)  ~~, ~~ \nabla_{\mu} \bar{h}^{\mu}_{\phantom{\mu}\nu} = 0 ~~,
\end{equation}
where we have defined
\begin{equation}
	\bar{h}^{(D)}_{\mu\nu} = h^{(D)}_{\mu\nu} - \frac{h^{(D)}}{2} \eta_{\mu\nu}~~,~~ h^{(D)} = \eta^{\mu\nu} h^{(D)}_{\mu\nu} ~~,
\end{equation}
and the Laplacian operator is taken along transverse directions to the worldvolume. The direction cosine $r_{\perp} = r\cos\theta$ is transverse to the direction along which the brane is bent. It is convenient to exhibit the explicit $r$- and $\theta$-dependence of the the asymptotic form of the dipole contributions, thus we define
\begin{equation} \label{ffs}
	 h^{(D)}_{ab} = f^{(D)}_{ab} \cos\theta \frac{r_0^{n+2}}{r^{n+1}} ~~,~~ h^{(D)}_{rr} = f^{(D)}_{rr} \cos\theta \frac{r_0^{n+2}}{r^{n+1}}~~,~~h^{(D)}_{ij} = r^2 g_{ij} f^{(D)}_{\Omega\Omega} \cos\theta \frac{r_0^{n+2}}{r^{n+1}} ~~,
\end{equation}
where $f^{(D)}_{\mu\nu}$ are the asymptotic metric coefficients which do not depend on $r$ neither on $\theta$\footnote{Here $f^{(D)}_{\Omega\Omega}$ is the same function for all transverse sphere indices.}. With this definition, the transverse gauge condition gives rise to the constraint
\begin{equation} \label{eqn:transversegaugecond}
	\eta^{ab} f^{(D)}_{ab} + f^{(D)}_{rr} + (n-1)  f^{(D)}_{\Omega\Omega} = 0 ~~,
\end{equation}
and hence one obtains
\begin{equation} \label{eqn:tracegaugecond}
	h^{(D)} =  2 f^{(D)}_{\Omega\Omega}   \cos\theta \frac{r_0^{n+2}}{r^{n+1}} ~~.
\end{equation}
The dipole contributions to the metric are therefore given by\footnote{Note that here we have defined $d_{ab} = \frac{\Omega_{(n+1)}r_0^n}{16\pi G} r^2_0 \hat{d}_{ab}$ and omitted the transverse index $r_\perp$ from ${d_{ab}}^{r_\perp}$ since, according to the analysis of \cite{Camps:2012hw}, also valid for the case at hand, perturbations in each direction $r_\perp$ decouple from each other to first order in the derivative expansion.}
\begin{equation} \label{eqn:dmunuCoeff}
	\hat{d}_{ab} = \bar{f}^{(D)}_{ab} = f^{(D)}_{ab} - f^{(D)}_{\Omega\Omega} \eta_{ab} ~~,
\end{equation} 
and hence the Young modulus $\tilde Y^{abcd}$ can then be obtained via Eq.\eqref{eqn:YMdef}.

\subsubsection*{Measuring the piezoelectric moduli} 

The procedure for obtaining the piezoelectric moduli follows a similar logic. Using the linearized version of Eq.~\eqref{eqn:gaugefieldeom} together with the expansion given by Eq.~\eqref{cpd} and corresponding decomposition (see Eq.~\eqref{decomp0} and Eq.~\eqref{decomp1_1}), one finds that the gauge field satisfies
\begin{equation} \label{Ar0}
	\nabla^2_{\perp} A^{(D)}_{\mu_1 \ldots \mu_{q+1}} = 16\pi G {p_{\mu_1 \ldots \mu_{q+1}}}^{r_{\perp}}\partial_{r_{\perp}} \delta^{(n+2)}(r) ~~, ~~ \nabla_{\mu} A^{\mu\nu_1 \ldots \nu_{q}} = 0 ~~,
\end{equation}
where it has been assumed that the dilaton vanishes at infinity. Again, it is convenient to write the asymptotic gauge field coefficients as
\begin{equation} \label{Ar}
	A^{(D)}_{\mu_1 \ldots \mu_{q+1}} = a^{(D)}_{\mu_1 \ldots \mu_{q+1}} \thinspace  \cos\theta \frac{r_0^{n+2}}{r^{n+1}} ~~.
\end{equation}
The electric dipole moment \eqref{eq:piezo} follows from Eqs.~\eqref{Ar0}-\eqref{Ar} leading to the simple relation\footnote{Note that here we have defined $p_{a_1 \ldots a_{q+1}} = \frac{\Omega_{(n+1)}r_0^n}{16\pi G} r^2_0 \hat{p}_{a_1 \ldots a_{q+1}}$ and again omitted the transverse index $r_\perp$ from ${p_{a_1 \ldots a_{q+1}}}^{r_\perp}$ in parallel with our definition of ${d_{ab}}^{r_\perp}$ (see footnote 8).}
\begin{equation} \label{eqn:dipoleCurrent}
	\hat{p}_{a_1 \ldots a_{q+1}} = a^{(D)}_{a_1 \ldots a_{q+1}} ~~.
\end{equation}
The piezoelectric moduli $\tilde\kappa^{a1_...a_{q+1}bc}$ can then be extracted from \eqref{eqn:dipoleCurrent} via Eq.~\eqref{eq:piezo2}. This concludes our review of how the response coefficients are obtained from the field content of bent black brane solutions far away from the brane horizon. We shall now turn to the construction of actual solutions and provide their response coefficients as examples of this procedure.

\subsection{Solution generating techniques} \label{sec:solutions}

As mentioned in the previous section, we consider the generalized Gibbons-Maeda solutions of the action \eqref{eqn:action} and obtain bent versions for a subset of these. These solutions consist of dilatonic black $p$-branes with smeared electric $q$-charge. Here we present the leading order solution for which the metric is given by
\begin{equation} \label{eqn:GMmetric}
	\text{d}s^2 = h^{-A} \left( - f \text{d}t^2 + \text{d}\vec{y} \right) + h^B\left( f^{-1} \text{d}r^2 + r^2\text{d}\Omega^2_{(n+1)} + \text{d}\vec{z} \right) ~~,
\end{equation}
where $\vec{y}$ labels the $q$ directions in which the gauge field has non-zero components and $\vec{z}$ labels the remaining $p-q$ smeared directions. The two harmonic functions entering \eqref{eqn:GMmetric} are
\begin{equation}
	f(r) = 1 - \frac{r_0^n}{r^n} \quad \text{and} \quad h(r) = 1 + \frac{r_0^n}{r^n} \sinh^2\alpha ~~,
\end{equation}
where the two parameters $r_0$ and $\alpha$ are the horizon radius and charge parameter respectively, which are related to the temperature and chemical potential of the solution. The gauge field in turn is given by
\begin{equation}
	A_{[q+1]} = -\sqrt{N} \frac{r_0^n}{r^n h(r)} \cosh\alpha \sinh\alpha \; \text{d}t \wedge \text{d}y^1 \wedge \ldots \wedge \text{d}y^q ~~,
\end{equation}
while the dilaton reads
\begin{equation} \label{eqn:GMdilaton}
	\phi = \frac{1}{2} N a \log h(r) ~~,
\end{equation}
where $N = A+B$. Finally, the exponents $A$ and $B$ are constant numbers depending on $p,q,n$ and the dilaton coupling $a$. We will provide these for the particular subset of solutions considered below, all of which satisfy $N=1$. From the solution \eqref{eqn:GMmetric}-\eqref{eqn:GMdilaton} it is straight forward to obtain the monopole corrections $h_{\mu\nu}^{(M)}$ and $A^{(M)}_{\mu_1...\mu_{q+1}}$, appearing in the decompositions \eqref{eqn:decomp} and \eqref{eqn:decompB} respectively, via Eqs.~\eqref{eqn:einsteineqn}-\eqref{eqn:gaugefieldeom}. 

\subsubsection*{Classes of bent metrics}
In order to obtain the dipole corrections $h_{\mu\nu}^{(D)}$ and $A^{(D)}_{\mu_1...\mu_{q+1}}$ for a subset of the solutions \eqref{eqn:GMmetric}-\eqref{eqn:GMdilaton} we use different solution generating techniques. These techniques allows us to construct bent black branes with smeared $q$-brane charge and Kaluza-Klein dilaton coupling. To generate these charged geometries we take as seed solution the elastically perturbed neutral black brane obtained in Ref.~\cite{Camps:2012hw}, which is reviewed in detail in App.~\ref{app:neutral}.

The first class of solutions that we consider consists of black dilatonic $p$-branes with a single Maxwell gauge field. This class is constructed by uplifting the seed solution with $m+1$ additional flat directions. The resulting metric is then boosted along the time direction and one of the uplifted directions and followed by a Kaluza-Klein reduction along that particular uplifted and boosted direction. In this way we obtain $p$-brane solutions carrying Maxwell charge ($q=0$). All the brane directions lie along the directions labelled by $\vec{z}$ which were introduced in \eqref{eqn:GMmetric}. The extra $m$ directions appearing as a byproduct of the uplift remain flat directions while the other worldvolume directions are now bent. The resulting solution will therefore be characterized by an isotropic stress-energy tensor. The details of this construction are presented in App.~\ref{sec:class1}.

The second class of solutions we consider are solutions to type II string theory in $D=10$ dimensions where we can use T-duality in order to generate higher-form gauge fields. The action describing this theory is given in Eq.~\eqref{eqn::SUGRAaction}. The solution generating technique works in the following way. Starting with a $p$-brane carrying $0$-brane charge, one can perform successive T-duality transformations on the $m$ flat directions leading to a $p$-brane with $q$-brane charge, where $q=m$. The effect of this transformation is to introduce higher-form fields and to unsmear the $m$ flat directions. In practice, this transforms the $m$ directions originally included in $\vec{z}$ into $m$ directions now included in $\vec{y}$. We thus end up with $Dq$-brane solutions smeared in $(p-q)$-directions constrained by the condition $n+p=7$ with $n \geq 1$. These solutions are characterized by an anisotropic stress-energy tensor to leading order. The details of this construction are presented in App.~\ref{sec:class2}.

It should be mentioned that both classes of generated solutions are valid for $n\geq 1$, but in order to measure the response coefficients one must require that $n \geq 3$ such that self-gravitational interactions are sub-leading with respect to the fine structure corrections \cite{Armas:2011uf}. Also, it is crucial to point out that we measure the response coefficients under the assumption that the extrinsic curvature components satisfy $K_{ta}=0$ for all $a$. If this was not the case the solution generating technique that we use here would introduce a background gauge field and a non-zero background dilaton which we did not consider in measurement procedure outlined in the beginning of Sec.~\ref{sec:measurement}.

\subsection{Worldvolume stress-energy tensor and thermodynamics}
Here we present the thermodynamic quantities characterizing the worldvolume stress-energy tensor $T^{ab}_{(0)}$ and worldvolume electric current $J^{a_1...a_{q+1}}_{(0)}$ of the solutions generated using the methods outlined in the previous section. The worldvolume stress-energy tensor takes the form\footnote{Notice that this form of the wolrdvolume stress-energy tensor and current is certainly not the most general form for the stress-energy tensor and current of the Gibbons-Maeda family of solutions with $q<p$ charge, however, for the cases $q=0,1$ the form presented here is the most general form \cite{Caldarelli:2010xz}.}
\begin{equation} \label{eqn:stresstensorq2}
T^{ab}_{(0)}=\varepsilon u^{a}u^{b} 
+ P_{\perp}\left(\gamma^{ab}+u^{a}u^{b}-\sum_{i=1}^{q}v^{a}_{(i)}v^{b}_{(i)}\right)
+ P_{\parallel}\sum_{i=1}^{q}v^{a}_{(i)}v^{b}_{(i)}~~,
\end{equation}
while the worldvolume electric current reads
\begin{equation} \label{eqn:current}
J^{a_{1}...a_{q+1}}_{(0)}=(q+1)!\mathcal{Q}u^{[a_1} v^{a_2}_{(1)}...v^{a_{q+1}]}_{(q)} ~~.
\end{equation}
Note that the form presented here is the same as that obtained to leading order in the expansion. Indeed, as noted in \cite{Emparan:2007wm,Camps:2012hw,Armas:2012ac} for elastically perturbed black branes there are no corrections to the worldvolume stress-energy tensor $T^{ab}_{(0)}$ to order $\mathcal{O}(\tilde\varepsilon)$. This fact is supported by a general analysis of the effective action for stationary black holes \cite{Armas:2013hsa} which applies to the cases studied here. Therefore, the thermodynamic quantities entering \eqref{eqn:stresstensorq2}-\eqref{eqn:current} do not suffer corrections to this order. Thus, the energy density $\varepsilon$, the pressure $P_{\parallel}$ in the $q$-directions, and the pressure $P_{\perp}$ in the $p-q$ remaining directions of the worlvolume are given by the leading order results \cite{Caldarelli:2010xz}
\begin{equation}
\begin{split}
	\varepsilon &= \frac{\Omega_{(n+1)}}{16\pi G} r_0^n \left( n + 1 + nN \sinh^2 \alpha \right) \,, \\
	P_{\parallel} &= -\frac{\Omega_{(n+1)}}{16\pi G} r_0^n \left( 1 + nN\sinh^2\alpha \right) \,, \quad
	P_{\perp} = - \frac{\Omega_{(n+1)}}{16\pi G} r_0^n ~~,
\end{split}
\end{equation}
where the various quantities are parameterized in terms of a charge parameter $\alpha$ and the horizon radius $r_0$. Furthermore, the local temperature $\mathcal{T}$, the local entropy density $s$, the local charge density $\mathcal{Q}$ and the local chemical potential $\Phi$ are given by
\begin{equation} \label{eqn:GMthermo}
\begin{split}
	\mathcal{T} &= \frac{n}{4\pi r_0 (\cosh\alpha)^N}, \quad 
	s = \frac{\Omega_{(n+1)}}{4G} r_0^{n+1} (\cosh\alpha)^N \,, \\
	\mathcal{Q} &= \frac{\Omega_{(n+1)}}{16\pi G} n \sqrt{N} r_0^n \cosh\alpha \sinh\alpha \,, \quad
	\Phi = \sqrt{N} \tanh\alpha ~~.
\end{split}
\end{equation}
Note again that for all the solutions constructed in this paper we have that $N=1$.

\subsection{Black branes carrying Maxwell charge} \label{sec:branesq0}

In this section we present the response coefficients for the first class of solutions described in Sec.~\ref{sec:solutions} consisting of dilatonic black $p$-branes with a single Maxwell gauge field. The details of this construction are presented in App.~\ref{sec:class1}. For the $q=0$ case the leading order worldvolume stress-energy tensor \eqref{eqn:stresstensorq2} is isotropic
\begin{equation} \label{eqn::GMstressq0}
	T^{ab}_{(0)} = \frac{\Omega_{(n+1)}}{16\pi G} r_0^n \left( n(1+N \sinh^2\alpha) u^a u^b - \gamma^{ab}  \right) ~~,
\end{equation}
while the leading order electric current is simply
\beq
J^{a}_{(0)}=\mathcal{Q}u^{a}~~.
\eeq
To first order in the perturbation expansion, solving Einstein equations for the perturbed metric such that the horizon remains regular requires solving the leading order blackfold equations \cite{Emparan:2007wm,Camps:2012hw} obtained by setting $d^{ab\rho}=0$ in Eqs.~\eqref{st11},\eqref{st12}. For a stress-energy tensor of the form \eqref{eqn::GMstressq0}, this implies that the following equation of motion must be satisfied
\begin{equation}
	n(1+N \sinh^2\alpha) u^a u^b K_{ab}^{\phantom{ab}i} = K^i ~~,
\end{equation}
where $K^i\equiv \gamma^{ab}{K_{ab}}^{i}$ is the mean extrinsic curvature vector. For example, by only considering perturbations along a single direction $\hat{i}$, i.e., an extrinsic curvature tensor of the form $K_{ab}^{\phantom{ab}\hat{i}} = \text{diag}(0,-1/R,0,\ldots)$ one finds the leading order critical boost $u_a = [\cosh\beta, \sinh\beta, 0, \ldots]$ where \cite{Caldarelli:2010xz}
\begin{equation} \label{eqn:critboost}
	\sinh^2\beta = \frac{1}{n(1+N \sinh^2\alpha)} ~~.
\end{equation}
The solution to the intrinsic equation \eqref{st11}, generically for $q=0$, is obtained by requiring stationarity of the overall configuration $u^{a}=\textbf{k}^{a}/\textbf{k}$ and setting the global horizon temperature $T$ and global horizon chemical potential $\Phi_H$ such that $T = |\mathbf{k}|\mathcal{T}$ and $\Phi_H = |\mathbf{k}| \Phi$ \cite{Caldarelli:2010xz}. Using these relations we can express the solution parameters $r_0$ and $\alpha$ in terms of the global quantities using the thermodynamic quantities given in \eqref{eqn:GMthermo} such that
\begin{equation} \label{eqn:q0intrinsicsol}
	r_0 = \frac{n}{4\pi T} |\mathbf{k}| \left( 1 - \frac{\Phi_H^2}{|\mathbf{k}|^2 } \right)^{\frac{1}{2}} ~~,~~
	\tanh\alpha = \frac{\Phi_H}{|\mathbf{k}| } ~~.
\end{equation} 
The solutions constructed in App.~\ref{sec:class1} are automatically stationary due to the stationarity of the neutral seed solution. The stress-energy tensor components of the solution to order $\mathcal{O}(\tilde\varepsilon)$ are
\begin{equation} \label{eqn:q0stress}
	T^{ab}_{(0)} = \frac{\Omega_{(n+1)}}{16\pi G} r_0^n \left( n \cosh^2\alpha u^a u^b - \eta^{ab}  \right) ~~, \quad
	T^{y_i y_j}_{(0)} = P_{\perp} \delta^{y_i y_j} ~~,
\end{equation}
with $a=(t,z_i)$. This result agrees with the form \eqref{eqn::GMstressq0} by noting that $u^{y_{i}}=0$. For $p=1$ and $m=1$ this would correspond to a charged tube. 

\subsubsection*{Response coefficients}

The components of the Young modulus can be obtained from the bending moment acquired by the bent metric which is given in Eq.~\eqref{eqn:dipolestressq0} together with Eq.~\eqref{eqn:YMdef}. It takes the covariant form
\begin{equation} \label{eqn:YMq0}
\begin{split}
\tilde{Y}_{ab}^{\phantom{ab}cd}=\;&P_{\perp} r_0^2 \xi_{2}(n)\cosh^2\alpha \left[ \frac{3n+4}{n^2(n+2)} \eta_{ab}\eta^{cd} + \frac{1}{(n+2)\cosh^2\alpha}{\delta_{(a}}^{c}{\delta_{b)}}^{d}+2u_{(a}{\delta_{b)}}^{(c}u^{d)} \right] \\ 
&- \bar{k} \xi_2(n)r_0^2 \left[ T^{(0)}_{ab}\eta^{cd} + \eta_{ab}T_{(0)}^{cd} \right] ~~,
\end{split}
\end{equation}
where $\bar{k}$ is a dimensionless gauge parameter and the function $\xi_2(n)$ is given by
\begin{equation}
	\xi_2(n) = \frac{n \tan(\pi/n)}{\pi} \frac{\Gamma\left(\frac{n+1}{n}\right)^4}{\Gamma\left(\frac{n+2}{n}\right)^2} \,, \quad n \geq 3 \,.
\end{equation}
From Eq.~\eqref{YM_0} together with \eqref{eqn:q0intrinsicsol} we can obtain the associated non-vanishing $\lambda$-coefficients, which read
\begin{equation} \label{eqn:lambdacoeff0}
\begin{split}
	\lambda_1(\mathbf{k}; T, \Phi_H) &= \frac{\Omega_{(n+1)}}{16 \pi G} \xi_2(n) \left( \frac{n}{4\pi T} \right)^{n+2} |\mathbf{k}|^{n+2} \left( 1 - \frac{\Phi_H^2}{|\mathbf{k}|^2} \right)^\frac{n}{2} \left( \frac{3n+4}{2n^2(n+2)} - \bar{k} \left( 1 - \frac{\Phi_H^2}{|\mathbf{k}|^2} \right) \right) ~~, \\
	\lambda_2(\mathbf{k}; T, \Phi_H) &= \frac{\Omega_{(n+1)}}{16 \pi G} \xi_2(n) \left( \frac{n}{4\pi T} \right)^{n+2} |\mathbf{k}|^{n+2} \left( 1 - \frac{\Phi_H^2}{|\mathbf{k}|^2} \right)^{\frac{n}{2}+1} \frac{1}{2(n+2)} ~~, \\
	\lambda_3(\mathbf{k}; T, \Phi_H) &= \frac{\Omega_{(n+1)}}{16 \pi G} \xi_2(n) \left( \frac{n}{4\pi T} \right)^{n+2} |\mathbf{k}|^{n} \left( 1 - \frac{\Phi_H^2}{|\mathbf{k}|^2} \right)^{\frac{n}{2}} ~~, \\
	\lambda_4(\mathbf{k}; T, \Phi_H) &= \lambda_3(\mathbf{k}; T, \Phi_H) n \bar{k}~~.
\end{split}
\end{equation}
Therefore, from \eqref{eqn:lambdacoeff0} we only have 3 independent transport coefficients. Similarly, the components of the piezoelectric moduli can be obtained from the electric dipole moment given in Eq.~\eqref{eqn:dipolescurrentq0} together with Eq.~\eqref{eq:piezo}. When written in a covariant form it reads
\begin{equation} \label{eqn:pmoduliq0}
\tilde{\kappa}_{a}^{\phantom{a}bc}=-\xi_{2}(n)r_0^2 \left( \frac{\mathcal{Q}}{n} {\delta_{a}}^{(b}u^{c)} 
+\bar{k} J^{(0)}_{a}\eta^{bc} \right) ~~.
\end{equation}
Again, we can obtain the associated non-vanishing $\kappa$-coefficients using Eq.~\eqref{PM_0} and the relations \eqref{eqn:q0intrinsicsol} yielding
\begin{equation} \label{eqn:kappacoeff0}
\begin{split}
	\kappa_1(\mathbf{k}; T, \Phi_H) &= \frac{\Omega_{(n+1)}}{16 \pi G} \frac{\xi_2(n)}{2}  \left( \frac{n}{4\pi T} \right)^{n+2} \Phi_H |\mathbf{k}|^{n} \left( 1 - \frac{\Phi_H^2}{|\mathbf{k}|^2} \right)^{\frac{n}{2}} ~~, \\
	\kappa_3(\mathbf{k}; T, \Phi_H) &= \kappa_1(\mathbf{k}; T, \Phi_H) n \bar{k} ~~,
\end{split}
\end{equation}
and therefore only one of the response coefficients is independent. Note that some of the coefficients presented in \eqref{eqn:lambdacoeff0} and \eqref{eqn:kappacoeff0} are gauge dependent. The Young modulus \eqref{eqn:YMq0} and the piezoelectric moduli \eqref{eqn:pmoduliq0} obtained here agree with the results of the case $p=1$ studied in Ref.~\cite{Armas:2012ac} when using the map given in App.~\ref{app:blackstring}. We conclude that the bent black branes carrying Maxwell charge constructed in this paper are characterized by a total of 3+1=4 independent response coefficients.

\subsection{Smeared black D\texorpdfstring{$q$}{q}-branes}

In this section we specialize to black branes in type II string theory in $D=10$ and present the corrections to the stress-energy tensor \eqref{stpd} and current \eqref{cpd} as well as the response coefficients. This class of solutions consists of black $p$-branes carrying D$q$-charge and are solutions to the equations of motion that follow from the action \eqref{eqn::SUGRAaction}. The details of this construction are presented in App.~\ref{sec:class2}. The components of the monopole source of stress-energy tensor can be obtained from the solution given in Eq.~\eqref{eqn:Tdualsol} and read
\begin{equation} \label{eqn:GMstress1}	
	T^{ab}_{(0)} = \frac{\Omega_{(n+1)}}{16\pi G} r_0^n \left( n \cosh^2\alpha u^a u^b - \eta^{ab}  \right) ~~, \quad
	T^{y_i y_j}_{(0)} = P_{\parallel} \delta^{y_i y_j} ~~,
\end{equation}
with $a=(t,z_i)$. We thus see that the effect of the T-duality transformation is to unsmear the $\vec{y}$ directions, which can be easily realized when comparing the above stress-energy tensor with \eqref{eqn:q0stress}. For this particular class of solutions the stress-energy tensor given in \eqref{eqn:GMstress1} can be put into the form \eqref{eqn:stresstensorq2} by taking $u^{y_i} =0$ and noting that the $v_{y_i}^{(i)}$ vectors only take values in the $y$-directions, e.g., $v_{y_i}^{(2)}=(0,1,0,...,0)$. Similarly, the electric current can be put into the form \eqref{eqn:current}.

\subsubsection*{Branes carrying string charge}

In the case of $q=1$ the worldvolume stress-energy tensor \eqref{eqn:stresstensorq2} reduces to
\begin{equation} \label{eqn::GMstressq1}
	T^{ab}_{(0)} = \frac{\Omega_{(n+1}}{16\pi G} r_0^n \left( n u^a u^b - \gamma^{ab} - nN\sinh^2\alpha ( - u^au^b + v^a v^b ) \right) ~~,
\end{equation}
where we have omitted the index $(1)$ from the vector $v_{a}^{(1)}$. The leading order equilibrium condition for configurations with $N=1$ is obtained by solving Eq.~\eqref{st12} when $d^{ab\rho}=0$ such that
\begin{equation}
	n( u^a u^b \cosh^2\alpha - v^a v^b \sinh^2\alpha)  K_{ab}^{\phantom{ab}i} = K^i ~~,
\end{equation}
while the leading order solution to the intrinsic equation \eqref{st11}, in the case of $q=1$, is again obtained by the requirement of stationarity $u^{a}=\textbf{k}^{a}/\textbf{k}$ and by taking $T = |\mathbf{k}|\mathcal{T}$ and $\Phi_H = 2\pi |\bv| |\mathbf{k}| \Phi$ \cite{Caldarelli:2010xz}. The solution parameters $r_0$ and $\alpha$ expressed in terms of global quantities using the thermodynamic quantities \eqref{eqn:GMthermo} read
\begin{equation} \label{eqn:q1intrinsicsol}
	r_0 = \frac{n}{4\pi T} |\mathbf{k}| \left( 1 - \frac{1}{2\pi} \frac{\Phi_H^2}{(|\mathbf{k}| |\bv|)^2 } \right)^{\frac{1}{2}}~~,~~
	\tanh\alpha = \frac{1}{2\pi} \frac{\Phi_H}{|\mathbf{k}| |\bv| } ~~.
	\end{equation}
The components of the Young modulus can be obtained from the dipole contributions to the metric given in Eq.~\eqref{eqn:dipolestressqm}. This results in the same form as that obtained previously in \eqref{eqn:YMq0} but now with $T^{(0)}_{ab}$ given by Eq.~\eqref{eqn::GMstressq1}. The associated non-vanishing $\lambda$-coefficients now enjoy the intrinsic dynamics defined by the relations \eqref{eqn:q1intrinsicsol} and are given by 
\begin{align} \label{eqn:lambdacoeff1}
	\lambda_1(\mathbf{k}, \bv; T, \Phi_H) &= \frac{\Omega_{(n+1)}}{16 \pi G} \xi_2(n) \left( \frac{n}{4\pi T} \right)^{n+2} |\mathbf{k}|^{n+2} \left( 1 - \frac{1}{2\pi} \frac{\Phi_H}{|\mathbf{k}||\bv|} \right)^\frac{n}{2} \left( \frac{3n+4}{2n^2(n+2)} - \bar{k} \left( 1 - \frac{1}{2\pi} \frac{\Phi_H}{|\mathbf{k}||\bv|} \right) \right) ~~, \nonumber\\
	\lambda_2(\mathbf{k}, \bv; T, \Phi_H) &= \frac{\Omega_{(n+1)}}{16 \pi G} \xi_2(n) \left( \frac{n}{4\pi T} \right)^{n+2} |\mathbf{k}|^{n+2} \left( 1 - \frac{1}{2\pi} \frac{\Phi_H}{|\mathbf{k}||\bv|} \right)^{\frac{n}{2}+1} \frac{1}{2(n+2)} ~~, \nonumber\\
	\lambda_3(\mathbf{k}, \bv; T, \Phi_H) &= \frac{\Omega_{(n+1)}}{16 \pi G} \xi_2(n) \left( \frac{n}{4\pi T} \right)^{n+2} |\mathbf{k}|^{n} \left( 1 - \frac{1}{2\pi} \frac{\Phi_H}{|\mathbf{k}||\bv|} \right)^{\frac{n}{2}} ~~, \\
	\lambda_4(\mathbf{k}, \bv; T, \Phi_H) &= \lambda_3(\mathbf{k}, \bv; T, \Phi_H) n \bar{k}~~. \nonumber
\end{align}
The piezoelectric moduli is a natural generalization of the $q=0$ case given in \eqref{eqn:pmoduliq0}. It can be obtained from the electric dipole moment given in Eq.~\eqref{eqn:dipolescurrentqm} and takes the form
\begin{equation}
\tilde{\kappa}_{ab}^{\phantom{ab}cd} = - \xi_{2}(n)r_0^2 \left( 2\frac{\mathcal{Q}}{n} {\delta_{[a}}^{(c}v_{b]}u^{d)} 
+ \bar{k} J^{(0)}_{ab}\eta^{cd} \right) ~~,
\end{equation}
where $\tilde{\kappa}_{ab}^{\phantom{ab}cd}$ satisfies the property $\tilde{\kappa}_{ab}^{\phantom{ab}cd}=\tilde{\kappa}_{[ab]}^{\phantom{[ab]}(cd)}$. Finally, the associated non-vanishing $\kappa$-coefficients are given by
\begin{equation} \label{eqn:kappacoeff1}
\begin{split}
	\kappa_1(\mathbf{k}, \bv; T, \Phi_H) &= \frac{\Omega_{(n+1)}}{16 \pi G} \frac{\xi_2(n)}{2}  \left( \frac{n}{4\pi T} \right)^{n+2} \Phi_H |\mathbf{k}|^{n} \left( 1 - \frac{1}{2\pi} \frac{\Phi_H}{|\mathbf{k}||\bv|} \right)^{\frac{n}{2}} ~~, \\
	\kappa_3(\mathbf{k}, \bv; T, \Phi_H) &= \kappa_1(\mathbf{k}, \bv; T, \Phi_H) n \bar{k} ~~.
\end{split}
\end{equation}
Again notice that a subset of the coefficients presented in \eqref{eqn:lambdacoeff1} and \eqref{eqn:kappacoeff1} show gauge dependence. We conclude that these branes carrying string charge are characterized by a total of 3+1=4 independent response coefficients.

\subsubsection*{Branes charged under higher-form fields}

For the case $1<q<p$, one can again obtain the Young modulus from the bending moment given in Eq.~\eqref{eqn:dipolestressqm}. It will again lead to the expression written in Eq.~\eqref{eqn:YMq0} but now with $T_{(0)}^{ab}$ given by \eqref{eqn:stresstensorq2}. For the piezoelectric moduli, by means of Eq.~\eqref{eqn:dipolescurrentqm}, we find the natural generalization
\begin{equation}
\tilde{\kappa}_{ba_{1}...a_{q}}^{\phantom{ba_{1}...a_{q}}cd} = -\xi_{2}(n)r_0^2 \left( (q+1)!\frac{\mathcal{Q}}{n} {\delta_{[b}}^{(c}v_{a_1}^{(1)}...v_{a_{q}]}^{(q)}u^{d)} 
+\bar{k} J^{(0)}_{ba_{1}...a_{q}}\eta^{cd} \right) ~~,
\end{equation}
which satisfies the property $\tilde{\kappa}_{ba_{1}...a_{q}}^{\phantom{ba_{1}...a_{q}}cd}=\tilde{\kappa}_{[ba_{1}...a_{q}]}^{\phantom{[ba_{1}...a_{q}]}(cd)}$.

We thus find that the Young modulus and the piezoelectric moduli of all the strained charged black brane solutions considered in this paper can be parameterized by a total of 4 response coefficients. The fact that we find the same form for the response coefficients associated with the Young modulus is not surprising, since all the solutions obtained here are only perturbed along smeared directions.


\section{Discussion} \label{sec:discussion}

We have obtained the general form of the equations of motion to pole-dipole order of fluid branes carrying higher-form charge,
generalizing the results for neutral pole-dipole branes obtained in \cite{Vasilic:2007wp,Armas:2011uf}.
These results are important in understanding finite size effects when applying the blackfold approach \cite{Emparan:2009cs,Emparan:2009at} to charged branes in supergravity \cite{Emparan:2011hg,Caldarelli:2010xz,Grignani:2010xm}. 
While this has been our main motivation, these results may be of more general use in the study of charged extended objects
in other settings.  

By assuming linear response theory we have, following \cite{Armas:2011uf,Armas:2012ac}, subsequently proposed 
general forms of the relevant response coefficients, Young modulus and piezoelectric moduli, that characterize stationary bent charged (an)isotropic fluid branes. These results, together with the leading order corrections to the effective
action for stationary neutral fluid branes found recently in \cite{Armas:2013hsa},  constitute useful inputs towards the formulation 
of the general effective theory of thin elastic charged fluid branes. It would obviously be an interesting problem to find this effective theory and extend it to higher orders. In this connection we note that for the case
of $p$-branes with $q>1$ smeared charged, a complete characterization of stationary blackfold solutions has not been 
developed yet, which would be a first step towards constructing the general effective theory.
 Another important, but technically challenging, next step would be to obtain the metric of bent black branes
 to second order in the thin brane expansion. This would provide further clues to a more formal development of
 electroelasticity of black branes.  In that context, it would be interesting to investigate whether the linear response type
 relations of Eqs.~\eqref{eqn:YMdef}, \eqref{eq:piezo}, \eqref{eq:piezo2} and generalizations thereof, can be proven using general covariance and the laws of thermodynamics. 

To find explicit realizations in gravity of this electroelastic behavior, we have used solution generating techniques and neutral bent black brane geometries \cite{Emparan:2007wm,Camps:2012hw} as seed solutions, to construct large classes of bent charged black brane solutions, carrying Maxwell or higher-form charge smeared along worldvolume directions but transverse to the worldvolume current. In the former case
the branes are isotropic, while in the latter they are anisotropic. 
By measuring the bending moment and the electric dipole moment which these geometries acquire due to the strain, we
have then explicitly verified that these quantities are captured by classical electroelasticity theory.  In particular,
we found that the Young modulus and piezoelectric moduli of our strained charged black brane solutions are parameterized by a total of 3+1=4 response coefficients, both for the isotropic as well as anisotropic cases. 
While our branes provide an interesting
realization of electroelastic behavior of charged fluid branes, it is not surprising that they are characterized by just one
more response coefficient, as compared to bent neutral black branes. This is a consequence of the fact that  we obtain
them by a solution generating technique, which causes the branes to be bent only in the smeared directions.
It would therefore be very interesting to find more general bent black brane geometries, in which the bending also takes
place in the directions in which the brane is charged. This study will be presented elsewhere.

A particularly interesting special case of this would be to obtain, to first order, 
the metric of a bent D3-brane in type IIB string theory. This would allow to explore 
the physical interpretation of these response coefficients in the context of AdS/CFT.  
In this setting, we expect further finite thickness effects due to the coupling to the 5-form flux, namely an extra contribution to the dipole electric (magnetic) moment would appear which would allow us to measure electric (magnetic) susceptibilities. 
More generally, understanding the coupling of fluid branes to background fluxes and understanding polarization effects would
be interesting to pursue.  Moreover, examining the effect of Chern-Simons terms on the response coefficients computed 
in this paper, would also be relevant, in part due to the relation of these terms to the anomaly \cite{Banerjee:2008th,Erdmenger:2008rm,Son:2009tf} via the gauge/gravity correspondence. 
 
Further cases in gravity, including supergravity theories relevant to string theory, that would be worthwhile to study 
in the context of this paper would be spinning charged branes, allowing to see the responses corresponding to the magnetic moment discussed in  Sec.~\ref{sec:spincurrent}. In another direction one could generalize the analysis to  multi-charge brane configurations, for which interesting leading order blackfold
solutions  were discussed in Ref.~\cite{Emparan:2011hg}.  
We also note that examining the elastic corrections for thermal string probes   
\cite{Grignani:2010xm, Armas:2012bk, Grignani:2011mr}  is expected to shed further light on the physics
of these finite temperature objects, that where obtained using the blackfold method.

 Finally, it would be very interesting to examine whether the electroelastic behavior found here can provide clues towards the microscopics of black holes and branes.  The AdS/CFT context, already mentioned above,
 would probably be the most natural starting point for this, but more generally for the asymptotically flat black branes
considered in this paper, this holds the potential of providing valuable insights towards flat space holography. 
 Put another way, one may wonder whether there is a microscopic way to derive the type of response coefficients that
we have encountered in this work.


\section*{Acknowledgements}

We thank Joan Camps and Marko Vojinovic for useful discussions. JA thanks NBI for hospitality. 
The work of JA was partly funded by the Innovations- und Kooperationsprojekt C-13 of the Schweizerische Universit\"{a}tskonferenz (SUK/CRUS).
The work of NO is supported in part by the Danish National Research Foundation project \textbf{Black holes and their role in quantum gravity}.


\appendix


\section{Details on the derivation of the equations of motion} \label{app:eom}
In this appendix we give further details on the derivation of the equations of motion for $p$-branes carrying string charge ($q=1$).
\subsection{Pole-dipole branes carrying string charge}
Pole-dipole $p$-branes carrying string charge are characterized by an anti-symmetric current $J^{\mu\nu}$ of the form \eqref{cpd}. In order to solve Eq.~\eqref{ceom} we introduce an arbitrary tensor field $f_{\mu}(x^\alpha)$ of compact support and integrate \eqref{ceom} such that
\beq \label{intceom1}
\int d^{D}x\sqrt{-g}f_{\nu}(x^{\alpha})\nabla_{\mu}J^{\mu\nu}=0~~.
\eeq
We now decompose $f_{\mu}(x^\alpha)$ in parallel and orthogonal components to the worldvolume such that \cite{Vasilic:2007wp}
\begin{equation} \label{dddf}
\begin{split}
\nabla_{\lambda}f_{\mu}=f_{\mu\lambda}^{\perp}&+u^{a}_{\lambda}\nabla_{a}f_{\mu}~~,\\
\nabla_{(\rho}\nabla_{\lambda)}f_{\mu}=f_{\mu\lambda\rho}^{\perp}+2f_{\mu(\lambda a}^{\perp}u_{\rho)}^{a}+&f_{\mu a b}u^{a}_{\lambda}u^{b}_{\rho}~~,~~\nabla_{[\rho}\nabla_{\lambda]}f_{\mu}=\frac{1}{2}{R^{\sigma}}_{\mu\lambda\rho}f_{\sigma}~~.
\end{split}
\end{equation}
Explicit calculation using \eqref{dddf} and the projectors $u^{a}_{\lambda}$ and ${\perp^{\mu}}_{\lambda}$ leads to
\begin{equation} \label{fuab}
\begin{split}
f_{\mu ab}=\nabla_{(a}\nabla_{b)}f_{\mu}-f_{\mu\nu}^{\perp}\nabla_{a}u_{b}^{\nu}~~,~~f_{\mu \rho a}={\perp^{\nu}}_{\rho}\nabla_{a}f_{\mu\nu}^{\perp}+(\nabla_{a}u^{b}_{\rho})\nabla_{b}f_{u}+\frac{1}{2}{\perp^{\lambda}}_{\rho}u_{a}^{\nu}{R^{\sigma}}_{\mu\nu\lambda}f_{\sigma}~~.
\end{split}
\end{equation}
Eqs.~\eqref{dddf}-\eqref{fuab} indicate that on the worldvolume only the components $f_{\mu\nu\rho}^{\perp}$,~$f_{\mu\nu}^{\perp}$ and $f_{\mu}^{\perp}$ are mutually independent while on the boundary, as in Sec.~\ref{sec:dynamics}, we need to decompose $\nabla_{a}f_{\mu}$ such that \cite{Vasilic:2007wp}
\beq
\nabla_{a}f_\mu=\eta_a\nabla_\perp f_\mu +v^{\hat{a}}_{a}\nabla_{\hat{a}}f_{\mu}~~.
\eeq
On the brane boundary the components $f_{\mu\nu}^{\perp}$,~$\nabla_\perp f_\mu$ and $f_\mu$ are mutually independent. Given this, solving Eq.~\ref{intceom1} results in an equation with the following structure
\beq \label{eqstru}
\int_{\mathcal{W}_{p+1}}\sqrt{-\gamma}\left[Z^{\mu\nu\rho}f_{\mu\nu\rho}^{\perp}+Z^{\mu\nu}f_{\mu\nu}^{\perp}+Z^{\mu}f_\mu+\nabla_{a}\left(Z^{\mu\nu a}f_{\mu\nu}^{\perp}+Z^{\mu ab}\nabla_{b}f_\mu+Z^{\mu a}f_\mu\right)\right]=0~~. 
\eeq
This equation has the same structure as that obtained for the stress-energy tensor \eqref{stpd}. Requiring the first three terms to vanish independently leads to the equations of motion
\begin{equation} \label{eom11}
\perp_{\nu}^{\sigma}\perp_{\rho}^{\lambda}J^{\mu(\nu\rho)}_{(1)}=0~~,~~\perp_{\nu}^{\sigma}\left[J^{\mu\nu}_{(0)}-\nabla_{a}\left({\perp^{\nu}}_{\lambda}J^{\mu\lambda\rho}_{(1)}u_{\rho}^{a}+J^{\mu\rho\nu}_{(1)}u_{\rho}^{a}\right)\right]=0~~,
\end{equation}
\begin{equation} \label{eom21}
\nabla_{a}\left(J^{\mu\nu}_{(0)}u_{\nu}^{a}- 2J^{\mu(\nu\rho)}_{(1)}u^{[a}_{\rho}\nabla_{b}u^{b]}_{\nu} - \nabla_{b}\left(J^{\mu(\nu\rho)}_{(1)}\right)u^{b}_{\nu}u^{a}_{\rho} \right) -\left({\perp^{\lambda}}_{ \nu}J^{\sigma(\nu\rho)}_{(1)}{R^{\mu}}_{\sigma\rho\lambda}+\frac{1}{2}J^{\sigma\nu\rho}_{(1)}{R^{\mu}}_{\sigma\nu\rho}\right)=0~~.
\end{equation}
These equations have exactly the same form as those obtained for the stress-energy tensor \cite{Vasilic:2007wp}, the difference between the two is that now we are dealing with anti-symmetric tensors. Requiring the vanishing of the last three terms in Eq.~\eqref{eqstru} in terms of the mutually independent components leads to the boundary conditions
\beq \label{bound11}
{\perp^{\nu}}_{\lambda} J^{\mu(\lambda\rho)}_{(1)}u_{\rho}^{a}\eta_{a}|_{\partial\mathcal{W}_{p+1}}=0~~, ~~J^{\mu\lambda\rho}_{(1)}u_{\lambda}^{a}u_{\rho}^{b}\eta_{a}\eta_{b}|_{\partial\mathcal{W}_{p+1}}=0~~,
\eeq
\beq\label{bound21}
\left[\nabla_{\hat{a}}\left(J^{\mu(\lambda\rho)}_{(1)}u_{\lambda}^{a}u_{\rho}^{b}v^{\hat{a}}_{a}\eta_{b}\right)-\eta_a\left(J^{\mu\nu}_{(0)}u_{\nu}^{a}- 2J^{\mu(\nu\rho)}_{(1)}u^{[a}_{\rho}\nabla_{a}u^{b]}_{\nu} - \nabla_{b}\left(J^{\mu(\nu\rho)}_{(1)}\right)u^{b}_{\nu}u^{a}_{\rho}\right)\right]|_{\partial\mathcal{W}_{p+1}}=0~.
\eeq

We now want to recast Eqs.~\eqref{eom11}-\eqref{bound21} into a more convenient form. First we note that the first constraint in Eq.~\eqref{eom11} results in the decomposition of $J^{\mu\nu\rho}_{(1)}$ given in Eq.~\eqref{decomp1_1}. Now, as in the case of the stress-energy tensor \cite{Vasilic:2007wp} we introduce the analogous tensor structures
\beq \label{QM}
\mathcal{Q}^{\mu\nu a}=p^{ab[\mu}u_{b}^{\nu]}+J^{\mu\nu a}_{(1)}~~,~~\mathcal{M}^{\mu\nu a}=m^{a\mu\nu}-p^{ab[\mu}u_{b}^{\nu]}~~,
\eeq
where $\mathcal{Q}^{\mu\nu a}$ and $\mathcal{M}^{\mu\nu a}$ are both anti-symmetric in their space-time indices and furthermore $\mathcal{M}^{\mu\nu a}$ has the property $\mathcal{M}^{\mu\nu (a}u_{\nu}^{b)}=0$~. This means that the dipole correction to the current $J^{\mu\nu\rho}_{(1)}$ can be written as
\beq
J^{\mu\nu\rho}_{(1)}=2u^{[\mu}_{a}\mathcal{M}^{\nu]\rho a}+\mathcal{Q}^{\mu\nu a}u_{a}^{\rho}~~.
\eeq
Using these definitions in the second constraint given in \eqref{eom11} results in
\beq \label{eom11_b}
{\perp^{\sigma}}_{\nu}\left[J^{\mu\nu}_{(0)}-\nabla_{a}\left(\mathcal{Q}^{\mu\nu a}-\mathcal{M}^{\mu\nu a}\right)\right]=0~~.
\eeq
Now, making the most general decomposition of $J^{\mu\nu}_{(0)}$ which results in Eq.~\eqref{decomp1_00} and taking all the possible projections of Eq.~\eqref{eom11_b} leads to the relations
\beq \label{motion11_b}
J^{\rho a}_{\perp(1)}=u_{\mu}^{a}{\perp^{\rho}}_{\nu}\nabla_{a}(\mathcal{Q}^{\mu\nu a}-\mathcal{M}^{\mu\nu a})~~,~~J^{\sigma\rho}_{\perp(1)}=\perp^{\sigma}_{\mu}\perp^{\rho}_{\nu}\nabla_{a}\left(\mathcal{Q}^{\mu\nu a}-\mathcal{M}^{\mu\nu a}\right)~~.
\eeq
Note that, contrary to the equations of motion for the stress-energy tensor \cite{Vasilic:2007wp}, we only have two constraints. The third one, which in the case of \eqref{stpd} lead to the conservation of the spin current $j^{a\mu\nu}$, is non-existent here because both tensors introduced in \eqref{QM} are anti-symmetric. Finally, inserting the first relation in Eq.~\eqref{motion11_b} into Eq.~\eqref{eom21} we obtain the equation for current conservation
\beq
\nabla_a\left(\tilde{J}^{ab}u_{b}^{\mu}+u^{\mu}_{b}u^{a}_{\nu}u^{b}_{\rho}\nabla_{c}\mathcal{M}^{\nu\rho c}\right)=0~~,
\eeq
where we have defined the effective worldvolume current $\tilde{J}^{ab}=\tilde{J}^{[ab]}$ such that
\beq
\tilde{J}^{ab}=J^{ab}_{(0)}-u_{\mu}^{a}u_{\nu}^{b}\nabla_{c}\mathcal{Q}^{\mu\nu c}~.
\eeq
It is worth noticing that all the terms proportional to the Riemann tensor in Eq.~\eqref{eom21} have dropped out of the equations of motion as a consequence of the anti-symmetry of \eqref{cpd}. Moreover, at first sight, Eq.~\eqref{motion11_b} seems to contain two sets of independent equations obtained by projecting tangentially and orthogonally to the worldvolume. This is only apparent as the orthogonal projection of Eq.~\eqref{motion11_b},
\beq
\left(\tilde{J}^{ab}+u^{a}_{\nu}u^{b}_{\rho}\nabla_{c}\mathcal{M}^{\nu\rho c}\right){K_{ab}}^{\mu}=0~~,
\eeq
trivially vanishes due to the anti-symmetry of $\tilde{J}^{ab}$ and $\mathcal{M}^{\nu\rho c}$ and the symmetry of ${K_{ab}}^{\mu}$ in its two worldvolume indices. Taking the parallel projection and reintroducing $m^{a\mu\nu}$ and $p^{ab\mu}$ using \eqref{QM} leads to Eq.~\eqref{eom_2_q1}. Now, we turn our attention to the boundary conditions \eqref{bound11}-\eqref{bound21}. Introducing \eqref{QM} leads to
\beq
{\perp^{\nu}}_{\rho}\thinspace\mathcal{Q}^{\mu\rho a}\eta_a|_{\partial\mathcal{W}_{p+1}}=0~~,~~\left(\mathcal{M}^{\mu\rho a}-\mathcal{Q}^{\mu\rho a}\right)u_{\rho}^{b}\eta_{a}\eta_b|_{\partial\mathcal{W}_{p+1}}=0~~,
\eeq
\beq \label{bound21_b}
\left[\nabla_{\hat{a}}\left(J^{\hat{a}\hat{b}}_{(1)}v^{\mu}_{\hat{b}}\right)-\eta_a\left(\tilde{J}^{ab}u_{b}^{\mu}+u^{\mu}_{b}u^{a}_{\nu}u^{b}_{\rho}\nabla_{c}\mathcal{M}^{\nu\rho c}\right)\right]_{\partial\mathcal{W}_{p+1}}=0~~.
\eeq
Again, the orthogonal projection of Eq.~\eqref{bound21_b} vanishes and we are left with
\beq
\left[v^{b}_{\hat{b}}\nabla_{\hat{a}}J^{\hat{a}\hat{b}}_{(1)}-\eta_a\left(\tilde{J}^{ab}+u^{a}_{\nu}u^{b}_{\rho}\nabla_{c}\mathcal{M}^{\nu\rho c}\right)\right]_{\partial\mathcal{W}_{p+1}}=0~~.
\eeq
Reintroducing now $m^{a\mu\nu}$ and $p^{ab\mu}$ using \eqref{QM} leads to Eq.~\eqref{bound_2_q1}. As a final comment, we note that these equations are invariant under both extra symmetries and their transformation follows from Eq.~\eqref{extra1_q1} and Eq.~\eqref{extra2_q1} yielding
\beq
\delta_{1}\tilde J^{ab}=0~~,~~\delta_{1}\mathcal{M}^{\mu\nu a}=0~~,
\eeq
\beq
\delta_{2}\tilde J^{ab}=-J^{ab}_{(0)}u^{c}_{\rho}\nabla_{c}\tilde\varepsilon^{\rho}+J^{c[a}_{(0)}u^{b]}_{\rho}\nabla_{c}\tilde\varepsilon^{\rho}+\nabla_{c}\left(J^{ab}_{(0)}\tilde\varepsilon^{c}\right)~~,~~\delta_{2}\mathcal{M}^{\mu\nu a}=J^{ab}_{(0)}\tilde\varepsilon^{[\mu}_{\perp}u^{\nu]}_{b}~~,
\eeq
while the variations of $J^{\hat{a}\hat{b}}_{(1)}$ were given in Sec.~\ref{sec:stringcharge}.


\section{Elastically perturbed neutral black brane: a review} \label{app:neutral}

In this section we review the solution obtained in \cite{Camps:2012hw} and the adapted (Fermi normal) coordinates used to decouple the bending deformation along each orthogonal direction. We also provide some details on the calculations of the Young modulus and the relation between the solutions obtained here and the solution obtained in \cite{Armas:2011uf} for the bending of the black string.

\subsection{Extrinsic perturbations}

In order to study the extrinsic deformations of the worldvolume of a flat black $\tilde{p}$-brane one introduces a suitable set of adapted coordinates. A particular useful set of coordinates can be chosen such that the perturbations in directions orthogonal to the brane worldvolume decouple from each other. This is achieved by using Fermi normal coordinates. Since the extrinsic curvature of the worldvolume is the only first order derivative correction which characterizes the bending of the brane \cite{Armas:2013hsa}, it is therefore possible to rewrite the induced metric on the brane in terms of the extrinsic curvature tensor $K_{ab}^{\phantom{ab}i}$.

In these coordinates the perturbations along each of the transverse directions $y^i$ decouple from each other and one can consider the deformation in each normal direction separately. One can therefore limit the analysis to the study where $K_{ab}^{\phantom{ab}i}$ is non-zero along a single direction $i = \hat{i}$. Introducing a direction cosine, $y^{\hat{i}} = r\cos\theta$, the uniformly boosted flat black $\tilde{p}$-brane metric in the adapted coordinates is given by \cite{Camps:2012hw}
\begin{align} \label{eqn:neutralmetric}
  \text{d}s^2_D = & \left( \eta_{ab} - 2 K_{ab}^{\phantom{ab}\hat{i}} \, r \cos\theta + \frac{r_0^n}{r^n} \tilde{u}_a \tilde{u}_b \right)\text{d}\sigma^a\text{d}\sigma^b + f^{-1}\text{d}r^2 + r^2 \text{d}\theta^2 + r^2 \sin^2\theta \text{d}\Omega_{(n)}^2  \nonumber \\
  &+ h_{\mu\nu}(r,\theta) \text{d}x^{\mu}\text{d}x^{\nu} + \mathcal{O}(r^2/R^2) ~~,
\end{align}
with
  $f(r) = 1 - \frac{r_0^n}{r^n}$.
We can now drop the index on $\hat{i}$ without loss of generality. Since the corrections are of dipole nature one can parametrize the extrinsic perturbation functions according to $h_{\mu\nu}(r,\theta) = \cos\theta \hat{h}_{\mu\nu}(r)$ with
\begin{equation} \label{eqn:hhat}
\begin{split}
	\hat{h}_{ab}(r) &= K_{ab} \mathsf{h}_1(r) + \tilde{u}^c \tilde{u}_{(a}K_{b)c} \mathsf{h}_2(r) + K \tilde{u}_a \tilde{u}_b h_{\gamma}(r) ~~, \\
	\hat{h}_{rr}(r) &= K f(r)^{-1} h_r(r) ~~, \\
	\hat{h}_{\Omega\Omega}(r) &= K r^2 h_{\Omega}(r) ~~.
\end{split}
\end{equation}
The solution is invariant under the coordinate transformation
\begin{equation}
	r \rightarrow r +  K \cos\theta \gamma(r) ~~,~~
	\theta \rightarrow \theta + K \sin\theta \int^{x} dx \frac{\gamma(x)}{x^2 f(x)} ~~,
\end{equation}
for which $\mathsf{h}_1(r)$ and $\mathsf{h}_2(r)$ are invariant, but the remaining functions transform according to
\begin{equation}
\begin{split}
	h_{\gamma}(r) & \rightarrow h_{\gamma}(r) - n \frac{r_0^n}{r^{n+1}} \gamma(r) ~~, \\
	h_{r}(r) & \rightarrow h_{r}(r) +  2\gamma'(r) - n \frac{r_0^n}{r^{n+1}} \frac{\gamma(r)}{f(r)} ~~, \\
	h'_{\Omega}(r) & \rightarrow h'_{\Omega}(r) + 2 \frac{\gamma'(r)}{r} + 2 \frac{r_0^n}{r^{n+2}} \frac{\gamma(r)}{f(r)} ~~.
\end{split}
\end{equation}
This coordinate-gauge freedom can be eliminated by forming invariant functions and taking combinations of the metric perturbations, e.g.
\begin{equation} \label{eqn:gaugeinvfunc}
\begin{split}
	\mathsf{h}_{r}(r) &= h_{r}(r) + \frac{2}{n} r_0 \left( \frac{r^{n+1}}{r_0^{n+1}} h_{\gamma}(r) \right)' - \frac{h_{\gamma}(r)}{f(r)} ~~, \\
	\mathsf{h}'_{\Omega}(r) &= h'_{\Omega}(r) + \frac{2}{n}\frac{r_0}{r} \left( \frac{r^{n+1}}{r_0^{n+1}} h_{\gamma}(r) \right)' + \frac{2}{nr}  \frac{h_{\gamma}(r)}{f(r)} ~~.
\end{split}
\end{equation}
The perturbations can then be expressed in terms of four coordinate-gauge invariant functions for which the solution is
\begin{equation}
\begin{split}
	\mathsf{h}_1(r) &= 2r - AP_{1/n}\left(2\frac{r^n}{r_0^n}-1\right)~~, \\
	\mathsf{h}_2(r) &= -A\frac{r_0^n}{r^n}\left[ P_{1/n}\left(2\frac{r^n}{r_0^n}-1\right) + P_{-1/n}\left(2\frac{r^n}{r_0^n}-1\right) \right]~~, \\
	\mathsf{h}_r(r) &= \frac{n+1}{n^2 f(r)} \left[\left(\frac{n}{n+1} - 2\frac{r_0^n}{r^n}\right)(2r-\mathsf{h}_1)-\mathsf{h}_2\right]~~, \\
	\mathsf{h}'_{\Omega}(r) &= \frac{1}{n r f(r)} \left( 2r - \mathsf{h}_1 + \frac{n+2}{2n}\mathsf{h}_2 \right)~~,
\end{split}
\end{equation}
with the constant
\begin{equation}
	A = 2 r_0 \frac{ \Gamma\left[\frac{n+1}{n}\right]^2 }{ \Gamma\left[\frac{n+2}{n}\right] }~~.
\end{equation}
Here $P_{\pm 1/n}$ are Legendre polynomials. The solution is ensured to be regular on the horizon for any extrinsic perturbation that satisfy the leading order extrinsic blackfold equations \eqref{st12} with $d^{ab\rho}=0$. Since the stress-energy tensor \eqref{eqn:stresstensorq2} of the neutral black $\tilde{p}$-brane is
\begin{equation} \label{eqn:neutralST}
	T^{ab}_{(0)} = \frac{\Omega_{(n+1)}}{16\pi G} r_0^n \left(n \tilde{u}^a \tilde{u}^b - \eta^{ab} \right) ~~,
\end{equation}
this is equivalent to the condition
	$n \tilde{u}^a \tilde{u}^b K_{ab} = K$.


\subsubsection*{Large $r$-asymptotics}

In order to find the dipole corrections to the stress-energy tensor and the bending moment of the brane we focus on the large $r$-asymptotics of the solution. Below, we list these for the neutral $\tilde{p}$-branes with $n \geq 3$. Given the asymptotics of the Legendre polynomials $P_{\pm 1/n}$, one finds the asymptotics of the coordinate-gauge invariant functions to be
\begin{equation} \label{eqn:asympt1}
\begin{split}
	\mathsf{h}_1(r) &= \frac{1}{n} \frac{r_0^n}{r^{n-1}} - \frac{\xi_2(n)}{n+2} \frac{r_0^{n + 2}}{r^{n + 1}} + \mathcal{O}(r^{-(n+2)}) ~~, \\
	\mathsf{h}_2(r) &= -2 \frac{r_0^n}{r^{n - 1}} - 2 \xi_2(n) \frac{r_0^{n + 2}}{r^{n + 1}} + \mathcal{O}(r^{-(n+2)}) ~~, \\
	\mathsf{h}_r(r) &= \frac{2}{n} r - \frac{3}{n^2}\frac{r_0^n}{r^{n - 1}} + \frac{4+7n+2n^2}{n^2(n+2)} \xi_2(n)  \frac{r_0^{n + 2}}{r^{n + 1}} + \mathcal{O}(r^{-(n+2)}) ~~, \\
	\mathsf{h}_{\Omega}(r) &= \frac{2}{n}r - \frac{n-3}{n^2(n-1)}\frac{r_0^n}{r^{n-1}} - \frac{4 + 3n + n^2}{n^2(n+2)(n+1)} \xi_2(n)  \frac{r_0^{n+2}}{r^{n+1}} + \mathcal{O}(r^{-(n+2)})~~. \\
\end{split}
\end{equation}
where\footnote{The function $\xi(n)$ given in \cite{Armas:2011uf} is related to $\xi_2(n)$ via $\xi(n) = \frac{n+1}{n^2(n+2)} \xi_2(n)$.}
\begin{equation}
	\xi_2(n) = \frac{ \Gamma\left[\frac{n - 2}{n}\right] \Gamma\left[\frac{n + 1}{n}\right]^2} {\Gamma\left[\frac{n + 2}{n}\right]\Gamma\left[\frac{n - 1}{n}\right]^2} = \frac{n \tan(\pi/n)}{4\pi r_0^2} A^2 ~~.
\end{equation}

\subsubsection*{Choosing a suitable gauge}

To find the actual large $r$-asymptotics of the metric one has to settle upon a gauge by choosing $\gamma(r)$. This choice is of course only for convenience, since it will not affect the actual response coefficients. The coordinate-gauge invariant functions are related to the metric perturbations via Eq.~\eqref{eqn:gaugeinvfunc}. Let us parameterize the asymptotics of the non-gauge invariant function $h_{\gamma}$ by
\begin{equation}
	h_{\gamma}(r) = b_0  r + b_1 \frac{r_0^n}{r^{n-1}} + k_1 \frac{r_0^{n+2}}{r^{n+1}} + b_4 \frac{r_0^{2n}}{r^{2n-1}} + k_2 \frac{r_0^{2n+2}}{r^{2n+1}} + \mathcal{O}\left(r^{-2n-2}\right) ~~,
\end{equation}
were the coefficients $b_0, b_1$ and $b_4$ are in principle determined by matching the asymptotics with the boundary conditions given in the overlap region \cite{Emparan:2007wm}. We are however free to choose a gauge where $b_0 = 0$ and $b_1 = \frac{1}{2}$. This leads to,
\begin{align} \label{eqn:asympt2}
	h_r(r) &= \left[ \frac{n^2 - 6 + (4 n^2 - 8 n)b_4 }{2 n^2} \right] \frac{r_0^n}{r^{n-1}} + \left[ (k_1 + 2k_2) + \frac{4+7n+2n^2}{n^2(n+2)} \xi_2(n) \right]  \frac{r_0^{n+2}}{r^{n+1}} + \mathcal{O}\left(r^{-n-2}\right)~~, \nonumber \\
	h_{\Omega}(r) &= \left[ \frac{3 - 2b_4(n^2 - 2 n)}{n^2(n-1)} \right] \frac{r_0^n}{r^{n-1}} + \left[ \frac{2(k_1 - nk_2)}{n(n+1)} + \frac{4 + 3n + n^2}{n^2(n+2)(n+1)} \xi_2(n) \right] \frac{r_0^{n+2}}{r^{n+1}} + \mathcal{O}\left(r^{-n-2}\right) ~~.
\end{align}
With this choice we eliminate some of the leading order terms in $h_r$ and $h_{\Omega}$. Note that we only need the dipole terms of these expansions in order to determine the response coefficients.

\subsection{Measuring the Young modulus}

In this section we provide the details on how the Young modulus is obtained by applying the procedure outlined in Sec.~\ref{sec:ComputingYM} to the case at hand. The dipole contributions can be read off from Eq.~\eqref{eqn:asympt1} and Eq.~\eqref{eqn:asympt2} such that
\begin{equation} \label{eqn:neutraldipoleterms}
\begin{split}
	f^{(D)}_{ab}(r) &= K_{ab} \left( - \frac{\xi_2(n)}{n+2} \right) + \tilde{u}^c \tilde{u}_{(a}K_{b)c} \left( - 2 \xi_2(n) \right) + K \tilde{u}_a \tilde{u}_b k_1 ~~, \\
	f^{(D)}_{rr}(r) &= K \left[ k_1 + 2k_2 + \frac{4+7n+2n^2}{n^2(n+2)} \xi_2(n) \right] ~~, \\
	f^{(D)}_{\Omega\Omega}(r) &= K \left[ \frac{2(k_1 - nk_2)}{n(n+1)} + \frac{4 + 3n + n^2}{n^2(n+2)(n+1)} \xi_2(n) \right] ~~,
\end{split}
\end{equation}
where the coefficients $f^{(D)}_{ab}(r)$ where defined in Eq.~\eqref{ffs}. Since the Young modulus is obtained using Eq.~\eqref{eqn:YMdef}, one is interested in the bending moment given by Eq.~\eqref{eqn:dmunuCoeff}.
With the transverse gauge condition given by Eq.~\eqref{eqn:transversegaugecond} and a redefinition of $k_2$ such that
\begin{equation}\label{eqn::tildegauge}
	k_1 = \frac{2}{{1-n}} \left( \xi_2(n) + n k_2 \right) \quad \text{and} \quad
	-\tilde{k} \xi_2(n) -  \frac{(n+1)(n-4)}{n^2(n^2+n-2)}  \xi_2(n) = k_2 \left( \frac{2}{1-n} \right) ~~,
\end{equation}
one obtains the following form for the bending moment
\begin{equation} \label{eqn:neutralresponse}
	\hat{d}_{ab} = - \xi_2(n) \left[ \frac{1}{n+2} K_{ab} + 2 \tilde{u}^c \tilde{u}_{(a}K_{b)c} + \frac{4+3n}{n(n+2)} K \tilde{u}_a \tilde{u}_b + \tilde{k} \left[ K(n \tilde{u}_a \tilde{u}_b - \eta_{ab}) \right] \right] ~~.
\end{equation}
Then using the fact that $n u^a u^b K_{ab} = K$ and together with \eqref{eqn:YMdef} one finally obtains the Young modulus
\begin{equation}
\begin{split}
	\tilde{Y}_{ab}^{\phantom{ab}cd} =& -  \frac{n \tan(\pi/n)}{4\pi} A^2 \bigg[ \frac{\Omega_{(n+1)} r_0^n}{16\pi G} 
	\left( \frac{1}{n+2} \delta_{(a}^{\phantom{(a}c} \delta_{b)}^{\phantom{b)}d} + 2 \tilde{u}_{(a}\delta_{b)}^{\phantom{b)}(c}\tilde{u}^{d)} +	\frac{3n+4}{n+2} \tilde{u}_a \tilde{u}_b \tilde{u}^c \tilde{u}^d \right) \\
	& + \left[ T^{(0)}_{ab} \eta^{cd} + \eta_{ab} T_{(0)}^{cd} \right] \tilde{k} \bigg] ~~,
\end{split}
\end{equation}
in agreement with what was found in \cite{Camps:2012hw}\footnote{Note that the Young modulus presented in \cite{Camps:2012hw} was defined with the opposite sign compared to the one presented here.}.

\subsection{Relation to the black string} \label{app:blackstring}

In this section we provide the relation to the special case of $\tilde{p}=1$ originally obtained in \cite{Armas:2011uf}. For this particular example we have $K_{ab} = \text{diag}\left(0, -1/R\right)$. The leading order stress-energy tensor of the black string is given by \eqref{eqn:neutralST}. Inserting the extrinsic curvature and boost in Eq.~\eqref{st12} with $d^{ab\rho}=0$ we find the condition for regularity at the horizon to be $s_{\alpha}^2 = 1/n$, where we have parameterized the boost as 
\begin{equation}
	\tilde{u}_{a} = \left(c_{\alpha}, s_{\alpha}\right) = \left(\sqrt{\frac{n+1}{n}}, \frac{1}{\sqrt{n}} \right)~~.
\end{equation}
Given this parameterization we have that
\begin{equation}
	\tilde{u}^c \tilde{u}_{(a}K_{b)c} =
	\frac{1}{R}
	\left[	
	\begin{array}{cc}
		0 & \frac{1}{2} c_{\alpha} s_{\alpha} \\
		\frac{1}{2} c_{\alpha} s_{\alpha} & s^2_{\alpha}
	\end{array}
	\right] ~~.
\end{equation}
This can then be used together with Eq.~\eqref{eqn:neutralresponse} to find the coefficients,
\begin{equation}
\begin{split}
	\hat{d}_{tt} &= \frac{2 k_2 n^2 (n+2)^2 + (n+1)(4 + 3 n + 2n^2) \xi_2(n)}{n^2 (n^2 + n - 2)R} ~~, \\
	\hat{d}_{tz} &= \frac{\sqrt{n+1}(2n k_2 + (n+1)\xi_2(n))}{ n(n-1) R} ~~, \\
	\hat{d}_{zz} &= \frac{(n+1)(4 + 3n)\xi_2(n)}{n^2(n+2)R}  ~~.
\end{split}
\end{equation}
In order to compare to the original result for the black string \cite{Armas:2011uf} one can use the relation 
\begin{equation}
	\xi_2(n) = \frac{n^2(n+2)}{n+1} \xi(n)~~,
\end{equation}
and change the gauge by
\begin{equation} \label{eqn:gaugerelation}
\quad k_2 = \frac{1}{2} ((1-n)\bar{k}_2 - n (n+2) \xi(n))~~,
\end{equation}
for which one finds the expected form
\begin{equation}
	\hat{d}_{tt} = - \frac{1}{R} \left( \bar{k}_2 (n + 2) + (n^2 + 3 n + 4) \xi(n) \right) ~~, \quad
	\hat{d}_{tz} = - \frac{\bar{k}_2}{R}  \sqrt{n+1} ~~, \quad
	\hat{d}_{zz} = \frac{1}{R} (3 n + 4) \xi(n) ~~.
\end{equation}
Finally, we note that $\bar{k}_2 = 0$ corresponds, via Eq.~\eqref{eqn:gaugerelation}, to
\begin{equation}
	k_2 = -\frac{n(n+2)}{2} \xi(n) = - \frac{n + 1}{2n} \xi_2(n) ~~,
\end{equation}
which by Eq.~\eqref{eqn::tildegauge} leads to 
\begin{equation}
	\tilde{k} = -\frac{(n+1)(n+4)}{n^2(n+2)} ~~.
\end{equation}

\section{Generating solutions and explicit form of metrics} \label{app:metrics}

In this appendix we provide the details on the construction of the two classes of solutions described in Sec.~\ref{sec:solutions} using the different solution generating techniques.

\subsection{Uplift-boost-reduce transformation} \label{sec:class1}

We generate charged solutions by using the solution generating technique consisting of applying an uplift-boost-reduce transformation to the neutral solution given in App.~\ref{app:neutral}. In this way we obtain the first order extrinsic perturbations to charged solutions. The end configuration will consist of dilatonic black $p$-brane metrics charged under a Maxwell gauge field with Kaluza-Klein dilaton coupling.

The first step in this construction is to uplift the $D$-dimensional seed solution given by Eq.~\eqref{eqn:neutralmetric} with $m+1$ additional flat directions,
\begin{equation} \label{eqn:upliftedmetric}
	\text{d}s^2_{d+1} = \text{d}s^2_{D} + \sum_{i=1}^{m} (\text{d}y_i)^2 + \text{d}x^2 ~~,
\end{equation}
where $d = \tilde{p}+m+n+3$. We denote the coordinates that span the original $\tilde{p}$-brane directions by $\sigma^a = (t, z^i)$ with $i=1,\ldots,\tilde{p}$. The additional flat directions are labeled by $y^i$ with $i=1,\ldots,m$. Lastly, we have separated the flat $x$-direction from the rest as it will serve as the isometry direction which we will perform the reduction over.

The second step is to apply an uniform boost $[c_{\kappa}, s_{\kappa}]$, with rapadity $\kappa$ along the $t$ and $x$-direction such that
\begin{equation*}
	g^{(d+1)}_{tt} = g_{tt} c^2_{\kappa} + s^2_{\kappa} ~~, \quad
	g^{(d+1)}_{xx} = g_{tt} s^2_{\kappa} + c^2_{\kappa} ~~, \quad
	g^{(d+1)}_{tx} = s_{\kappa}c_{\kappa}(g_{tt} + 1)  ~~, 
\end{equation*}
\begin{equation}
	g^{(d+1)}_{tz_i} = c_{\kappa} g_{tz_i} ~~, \quad
	g^{(d+1)}_{xz_i} = s_{\kappa} g_{tz_i} ~~,
\end{equation}
where $g^{(d+1)}$ is the boosted metric and $g$ is the metric given by \eqref{eqn:upliftedmetric}.

Finally, we can perform a reduction along the $x$-direction. The Einstein frame decomposition is given by
\begin{equation}
	\text{d}s^2_{(d+1)} = e^{2 \tilde{a} \phi} \text{d}s^2_{(d)} + e^{2(2-d)\tilde{a}\phi} ( \text{d}x + A_{\mu} \text{d}x^{\mu} )^2 ~~, \quad
	\tilde{a}^2 = \frac{1}{2(d-1)(d-2)} ~~,
\end{equation}
where $A_\mu$ is the gauge field and $\phi$ the dilaton. The Lagrangian density is decomposed according to
\begin{equation}
	\sqrt{-g_{(d+1)}} R_{(d+1)} = \sqrt{-g_{(d)}} \left( R_{(d)} - \frac{1}{2}(\partial\phi)^2 - \frac{1}{4}e^{-2(d-1)\tilde{a}\phi}
	F^2_{[2]} \right)~~,
\end{equation}
with $F_{[2]}=\text{d}A_{[1]}$. With this decomposition we find,
\begin{equation} \label{eqn:q0sol}
\begin{split}
	g^{(d)}_{\mu\nu} = e^{-2\tilde{a}\phi} \left( g^{(d+1)}_{\mu\nu} - \frac{ g^{(d+1)}_{\mu x}g^{(d+1)}_{\nu x} }{ g^{(d+1)}_{xx} } \right)  ~~,	\quad
	A_{\mu} = \frac{g^{(d+1)}_{x \mu}}{ g^{(d+1)}_{xx} } ~~, \quad
	e^{2(2-d)\tilde{a}\phi} = g^{(d+1)}_{xx} 
	~.
\end{split}
\end{equation}
Defining $p=\tilde{p}+m$ this provides us with the extrinsic perturbed solution of the black $p$-branes carrying $q=0$ charge in the presence of a dilaton. We thus have access to the explicit $1/R$ corrections to the fields. Note that only $\tilde{p}$ directions are extrinsically perturbed while the remaining $m$ directions remain flat. The rapidity $\kappa$ now takes the interpretation of a charge parameter.

\subsubsection*{Large \texorpdfstring{$r$}{r}-asymptotics}

We are interested in the large $r$-asymptotics of the solution and we provide them in terms of the large $r$-asymptotics of the neutral solution. We denote the boost velocities of the neutral solution by $\tilde{u}^{a}$. It is convenient to define the object
\begin{equation}
	H_{\mu\nu} = \cos\theta \left[ \hat{h}_{\mu\nu} - 2 r K_{\mu\nu} \right] ~~,
\end{equation}
with $\hat{h}_{\mu\nu}$ given by  Eq.~\eqref{eqn:hhat} together with Eq.~\eqref{eqn:asympt1} and Eq.~\eqref{eqn:asympt2}. Here $K_{\mu\nu}$ is the extrinsic curvature tensor of the neutral brane solution which is equal to the extrinsic curvature tensor of the generated solution provided $K_{ta} = 0$ for all $a$, which we assume in the following. The large $r$-asymptotic behavior of the dilaton is
\begin{equation} \label{eqn:ubr1}
\begin{split}
	e^{-2\tilde{a}\phi} &= 1 + \frac{s^2_{\kappa} \tilde{u}_t^2}{d-2} \frac{r_0^n}{r^n} +  \frac{s^2_{\kappa}}{d-2}  
	F_{tt}
	 \left( 1 - \frac{d-3}{d-2}\frac{r_0^n}{r^n} s^2_{\kappa} \tilde{u}^2 \right) + \mathcal{O}\left(\frac{r_0^{n+2}}{r^{n+2}}\right) ~~.
\end{split}
\end{equation}
The large $r$-asymptotics of the metric components is
\begin{equation} \label{eqn:ubr2}
\begin{split}
	g_{tt} &= \eta_{tt} + \left( 1 + \frac{d-3}{d-2} s^2_{\kappa} \right) \frac{r_0^n}{r^n} \tilde{u}_t^2 
	+ H_{tt} \left( 1 + \frac{d-3}{d-2} s^2_{\kappa} + \frac{r_0^n}{r^n} \mathcal{C}_{tt} \right) + \mathcal{O}\left(\frac{r_0^{n+2}}{r^{n+2}}\right) ~~, \\
	g_{z_i z_j} &= \eta_{ij} + \left( \tilde{u}_{i} \tilde{u}_{j} + \eta_{i j} \frac{s^2_{\kappa} \tilde{u}_t^2}{d-2} \right)\frac{r_0^n}{r^n} 
	+ \left( H_{i j} + \eta_{i j} \frac{s^2_{\kappa} F_{tt}}{d-2} \right)
	+ \frac{r_0^n}{r^n} \mathcal{C}_{z_i z_j} +  \mathcal{O}\left(\frac{r_0^{n+2}}{r^{n+2}}\right) ~~, \\
	g_{t z_i} &= c_{\kappa} \frac{r_0^n}{r^n} \tilde{u}_t \tilde{u}_i + H_{t i} c_{\kappa} - 2 c_{\kappa} s^2_{\kappa} \tilde{u}_t \tilde{u}_{(t} H_{i) t}   \frac{d-3}{d-2} \frac{r_0^n}{r^n} + \mathcal{O}\left(\frac{r_0^{n+2}}{r^{n+2}}\right) ~~, \\
	g_{y_i y_i} &= 1 + \frac{s^2_{\kappa} \tilde{u}_t^2}{d-2} \frac{r_0^n}{r^n} +  \frac{s^2_{\kappa}}{d-2} H_{tt}
	 \left( 1 - \frac{d-3}{d-2}\frac{r_0^n}{r^n} s^2_{\kappa} \tilde{u}_t^2 \right) + \mathcal{O}\left(\frac{r_0^{n+2}}{r^{n+2}}\right) ~~, \\
	g_{rr} &= 1 + \left( 1 + \frac{s^2_{\kappa} \tilde{u}_t^2}{d-2} \right) \frac{r_0^n}{r^n} + H_{rr} + \frac{s^2_{\kappa}}{d-2} H_{tt} + \frac{r_0^n}{r^n} \mathcal{C}_{rr} + \mathcal{O}\left(\frac{r_0^{n+2}}{r^{n+2}}\right) ~, \\
	g_{\Omega\Omega} &= g_{\xi_i \xi_j} \left( 1 + \frac{s^2_{\kappa} \tilde{u}_t^2}{d-2}\frac{r_0^n}{r^n} \right) + 
	H_{\Omega\Omega} + \frac{g_{\xi_i \xi_j}s_{\kappa}^2}{d-2} H_{tt} + \frac{r_0^n}{r^n} \mathcal{C}_{\Omega\Omega} + \mathcal{O}\left(\frac{r_0^{n+2}}{r^{n+2}}\right) ~~,
\end{split}
\end{equation}
where
\begin{equation} \label{eqn:ubr2coeff}
\begin{split}
	\mathcal{C}_{tt} &= s^2_{\kappa} \tilde{u}_t^2 \frac{d-3}{d-2} \left( \frac{s^2_{\kappa}}{d-2} - 2 c^2_{\kappa} \right) ~~, \\
	\mathcal{C}_{z_i z_j} &= \frac{s^2_{\kappa}}{d-2}  \left[ \tilde{u}^2_{t} H_{i j} + \left( \tilde{u}_{i} \tilde{u}_{j} -  \eta_{i j} \frac{d-3}{d-2} s^2_{\kappa} \tilde{u}_t^2 \right) H_{tt}  \right] -2 s^2_{\kappa} \tilde{u}_t \tilde{u}_{(i} H_{j) t} ~~, \\
	\mathcal{C}_{rr} &= \frac{s^2_{\kappa}}{d-2} \left[ \tilde{u}_t^2 H_{rr} + \left( 1 - \frac{d-3}{d-2} s^2_{\kappa} \tilde{u}_t^2 \right) H_{tt} \right] ~~, \\
	\mathcal{C}_{\Omega\Omega} &= \frac{s^2_{\kappa} \tilde{u}_t^2}{d-2} \left( H_{\Omega\Omega} - g_{\xi_i \xi_j} H_{tt} \frac{d-3}{d-2} s^2_{\kappa} \right) ~~,
\end{split}
\end{equation}
with $H_{ij} = H_{z_i z_j}$, $\eta_{ij} = \eta_{z_i z_j}$ and $\tilde{u}_i = \tilde{u}_{z_i}$. The components $g_{\xi_i \xi_j}$ constitute the metric of a $(n+1)$-sphere of radius $r$. For the gauge field one finds in terms of the parameters of the neutral solution
\begin{equation} \label{eqn:ubr3}
\begin{split}
	A_{t} &= s_{\kappa} c_{\kappa} \left[ \frac{r_0^n}{r^n} \tilde{u}_t^2 + H_{tt} \left( 1 - 2 s^2_{\kappa} \frac{r_0^n}{r^n} \tilde{u}_t^2 \right) \right] + \mathcal{O}\left(\frac{r_0^{n+2}}{r^{n+2}}\right)~~, \\
	A_{z_i} &= s_{\kappa} \left[ \frac{r_0^n}{r^n} \tilde{u}_t \tilde{u}_i + H_{t i} - 
	2 s^2_{\kappa} \frac{r_0^n}{r^n} \tilde{u}_t \tilde{u}_{(t} H_{i) t} \right] + \mathcal{O}\left(\frac{r_0^{n+2}}{r^{n+2}}\right)~~.
\end{split}
\end{equation}
With the large $r$-asymptotics we can now read off the dipole contributions and compute the response coefficients following the prescription given in Sec.~\ref{sec:solutions}. For this matter it is convenient to use the relation between the metric obtained here and the asymptotically flat generalized Gibbons-Maeda solutions.

\subsubsection*{Relation to the generalized Gibbons-Maeda solutions}

The solutions given by \eqref{eqn:q0sol} overlap with a subset of the asymptotically flat Gibbons-Maeda family presented in Sec.~\ref{sec:solutions}. The $\tilde{p}$-directions correspond to uniformly boosted directions with boost $u^{a}$ while the $m$-directions remain unaffected. It is possible to connect to this family of solutions by means of the identification of the dilaton coupling,
\begin{equation}
	a^2 = 4(d-1)^2\tilde{a}^2 = \frac{2(\tilde{p}+m+n+2)}{\tilde{p}+m+n+1} ~~.
\end{equation}
The leading order critical boost and charge parameter can be identified, e.g., from the components of the gauge field such that
\begin{equation} \label{eqn:GMrelations}
	\tilde{u}_{t} \cosh\kappa = u_{t} \cosh\alpha \quad \text{and} \quad
	\tilde{u}^i = u^i \cosh\alpha \quad \text{with} \quad 
	\sinh^2\alpha = \tilde{u}_t^2 \sinh^2\kappa ~~,
\end{equation}
where $\tilde{u}^{a}$ is the critical boost of the neutral solution while the normalization conditions $\tilde{u}^{a}\tilde{u}_{a}=u^{a}u_{a}=-1$ are satisfied. The critical boost is seen to be in agreement with Eq.~\eqref{eqn:critboost} for $N=1$. Finally, they correspond to solutions with
\begin{equation}
	A = \frac{d-3}{d-2}
	~~, ~~
	B = \frac{1}{d-2} ~~,
\end{equation}
and therefore $N=1$. 

\subsubsection*{Dipole terms}

The dipole contributions can be read off from the asymptotic expansion of the solution \eqref{eqn:ubr2} and \eqref{eqn:ubr3}. In terms of the dipole contributions of the seed solution one finds
\begin{equation}
\begin{split}
	\hat{f}^{(D)}_{tt} = \left( 1 + \frac{d-3}{d-2} s^2_{\kappa} \right) f^{(D)}_{tt} ~~, ~~
	\hat{f}^{(D)}_{t z_i} = c_{\kappa} f^{(D)}_{t z_i} ~~,~~
	\hat{f}^{(D)}_{z_i z_i} = \frac{1}{d-2} s^2_{\kappa} f^{(D)}_{tt} + f^{(D)}_{z_i z_i} ~~,
\end{split}
\end{equation}
\begin{equation*}
\begin{split}
	\hat{f}^{(D)}_{y_i y_i} = \frac{1}{d-2} s^2_{\kappa} f^{(D)}_{tt} ~~,~~
	\hat{f}^{(D)}_{r r} = \frac{1}{d-2} s^2_{\kappa} f^{(D)}_{tt} + f^{(D)}_{rr} ~~,~~
	\hat{f}^{(D)}_{\Omega \Omega} = \frac{1}{d-2} s^2_{\kappa} f^{(D)}_{tt} + f^{(D)}_{\Omega\Omega} ~~,
\end{split}
\end{equation*}
where $\hat{f}_{\mu\nu}$ denote the dipole coefficients of the charged solution (see Eq.~\eqref{ffs}). Notice that taking $\kappa$ to zero reproduces the result of the neutral seed solution.

Now in order to obtain the dipole contribution to the stress-energy tensor \eqref{stpd} we use the method outlined in Sec.~\ref{sec:solutions}. The transverse gauge condition given by Eq.~\eqref{eqn:transversegaugecond} naturally takes the form
\begin{equation} \label{eqn:ubrTraceCond1}
	(\eta^{ab} \hat{f}_{ab} + m \hat{f}_{yy})  + \hat{f}_{rr} + (n-1)  \hat{f}_{\Omega\Omega}  = 0~~.
\end{equation}
which ensures that relation obtained in \eqref{eqn::tildegauge} stays the same, that is, 
\begin{equation} \label{eqn:ubrTraceCond2}
	k_1 = \frac{2}{{1-n}} \left( \xi_2(n) + n k_2 \right) ~~,
\end{equation}
and similarly using Eq.~\eqref{eqn:tracegaugecond} one finds $\hat{f}^{(D)} = 2 \hat{f}_{\Omega\Omega}$. It is therefore possible to apply Eq.~\eqref{eqn:dmunuCoeff} to measure the bending moment. In terms of the neutral dipole coefficients one finds
\begin{equation} \label{eqn:dcompq0}
\begin{split}
	\hat{d}_{tt} = c^2_{\kappa} f^{(D)}_{tt} + f^{(D)}_{\Omega\Omega} ~~,
	\hat{d}_{t z_i} = c_{\kappa} f^{(D)}_{t z_i} ~~,
	\hat{d}_{z_i z_j} = f^{(D)}_{z_i z_j} - f^{(D)}_{\Omega\Omega} ~~,
	\hat{d}_{y_i y_j} = -f^{(D)}_{\Omega\Omega} ~~.
\end{split}
\end{equation}
Recall that the solution is not boosted along the $y^i$ directions and that these directions are flat, i.e., $K_{y_ia} = 0$ for all $a$ and $i$. Using the gauge choice given by Eq.~\eqref{eqn::tildegauge} such that
\begin{equation}
	k_{1}=-\frac{3n+4}{n(n+2)}\xi_{2}(n)-n\tilde k\xi_{2}(n) ~~,
\end{equation}
it is possible to see that in fact $f_{\Omega\Omega}^{(D)}=-\xi_{2}(n)\tilde k K$ and hence the dipole terms in those directions are indeed pure gauge.

Using the relations given in Eq.~\eqref{eqn:GMrelations} one can express the coefficients of the bending moment in terms of the generalized Gibbons-Maeda boost and charge parameters. Furthermore, it turns out that it is more natural to work in the gauge given by
\begin{equation}
	\tilde{k} = \bar{k} - \frac{3n+4}{n^2(n+2)} ~~,
\end{equation}
in the presence of charge, since the classical symmetries of the Young modulus will be manifestly apparent in this gauge. Suppressing the transverse index in the extrinsic curvature the
bending moment is found to be
\begin{equation}  \label{eqn:dipolestressq0}
\begin{split}
	\hat{d}_{ab} =\;
	& - \xi_2(n) \cosh^2\alpha \left[ \frac{ K_{ab}}{(n+2)\cosh^2\alpha} + 2 u^c u_{(a}K_{b)c} + \frac{3n+4}{n^2(n+2)} \eta_{ab} K \right]  \\ 
	& - \bar{k} \xi_2(n) \left[n \cosh^2\alpha \; u_au_b - \eta_{ab} \right]K   ~~,
\end{split}
\end{equation}
which, as mentioned in Sec.~\ref{sec:solutions}, is only valid under the assumption that all time components of the extrinsic curvature are zero, that is, $K_{ta} = 0$ for all $a$.

Similarly, for the gauge field, the non-vanishing dipole terms can be read off from the asymptotic expansion given in Eq.~\eqref{eqn:ubr3}
\begin{equation}
\begin{split}
	a^{(D)}_{t} = c_{\kappa}s_{\kappa} f^{(D)}_{tt} ~~,~~
	a^{(D)}_{z_i} = s_{\kappa} f^{(D)}_{t z_i} ~~.
	\end{split}
\end{equation}
Using Eq.~\eqref{eqn:dipoleCurrent} one can read off the electric dipole moment. Again, using the relations given in Eq.~\eqref{eqn:GMrelations} and the above assumptions we have that
\begin{equation} \label{eqn:dipolescurrentq0}
\begin{split}
	\hat{p}_{a} =\;
	& - \xi_2(n) \cosh\alpha \sinh\alpha \left[ u^c K_{ca} + \bar{k} u_a K \right] ~~.
\end{split}
\end{equation}
It is now possible to obtain the response coefficient from Eq.~\eqref{eqn:dipolestressq0} and Eq.~\eqref{eqn:dipolescurrentq0}, which are presented in Sec.~\ref{sec:solutions}.

\subsection{T-duality transformation} \label{sec:class2}

With the solutions given in the previous section it is possible to use T-duality on the residual $m$ isometries, if we impose $n+\tilde{p}+m=7$ and start from a solution with $m \geq 1$. In order to make contact with type II string theory in $D=10$ we consider the truncated effective action with zero NSNS 2-form $B$ field and only one R-R field\footnote{By using type IIB S-duality ($\phi \to -\phi$) it is not difficult to include the case of the NSNS 2-form field as well. In that way we can also obtain bent versions of smeared F-strings and NS5-branes. Note that for the case of $q=5$ we have $p=6$ and $n=1$, so while we can compute the bent geometries our results for the response coefficients are not valid since we need $n\geq 3$. For $q=1$ it is possible to have $n \geq 3$. In particular, we find that the response coefficients for the F1-string turn out to be the same as that of the D1-brane.}. Thus the configurations are solutions of the equations of motion that follow from the action
\begin{equation} \label{eqn::SUGRAaction}
	S = \int d^{10}x \sqrt{-g} \left[ R - \frac{1}{2}(\partial \phi)^2 - \frac{1}{2(q+2)!} e^{\frac{3-q}{2}\phi} H_{[q+2]}^2 \right] ~~.
\end{equation}
Let $z$ be an isometry direction, then in the Einstein frame, the T-duality transformation takes the form
\begin{equation}
\begin{split}
	g_{\mu\nu} &= e^{\frac{1}{8}\hat{\phi}} (\hat{g}_{zz})^{\frac{1}{4}}\left( \hat{g}_{\mu\nu} - \frac{\hat{g}_{\mu z} \hat{g}_{\nu z}}{\hat{g}_{zz}} \right) ~~,~~
	g_{zz} = e^{-\frac{7}{8}\hat{\phi}} (\hat{g}_{zz})^{-\frac{3}{4}}~~,~~ \\
	e^{2\phi} &= \frac{ e^{  \frac{3}{2} \hat{\phi}} } { \hat{g}_{zz} } ~~,~~
	A_{[q+2]} = A_{[q+1]} \wedge dz ~~,
\end{split}
\end{equation}
where the hatted quantities denote the fields before the transformation. The first T-duality transformation is applied to the solution given by \eqref{eqn:q0sol}\footnote{To relate our constructions with the black branes of supergravity we take $\hat{\phi} = -\phi$ with $\phi$ being the dilaton given in Eq.~\eqref{eqn:q0sol}. }. We apply the T-duality transformations in successive order to transform the $m$ flat directions and gain an $(m+1)$-form gauge field. Performing a recursive bookkeeping one finds the relation between the $m$'th T-duality transformation and the starting configuration to be
\begin{equation} \label{eqn:Tdualsol} 
	g_{\mu\nu} = e^{\frac{m}{6} \hat{\phi}} \hat{g}_{\mu\nu} ~~,~~
	g_{y_i y_j} = \delta_{ij} e^{\frac{m-7}{6} \hat{\phi}} ~~,~~
	\phi = \frac{3-m}{3} \hat{\phi} ~~.
\end{equation}
Here the $y^i$ directions come from the $m$ isometry directions and $\mu$ labels the remaining directions. Note that one can take $m=0$ and get the starting solution in which no $y^i$ directions are present. The solution now overlaps with the metric given in \eqref{eqn:GMmetric} where the $q=m$ directions included in $\vec{y}$ remain flat while the $\tilde{p}=p-q$ directions included in $\vec{z}$ are extrinsically perturbed.

\subsubsection*{Large \texorpdfstring{$r$}{r}-asymptotics}

The large $r$-asymptotics of the solution \eqref{eqn:Tdualsol} can be obtained from \eqref{eqn:ubr2}, \eqref{eqn:ubr2coeff} and \eqref{eqn:ubr3} by noting that
\begin{equation}
	A = \frac{1}{d-2} \rightarrow \frac{q+1}{8} ~~,~~ \text{and} \quad B = \frac{d-3}{d-2} \rightarrow \frac{7-q}{8}~~.
\end{equation}
The value $N=1$ is therefore also preserved under the transformation. The isometry directions naturally differ from the rest and leads to
\begin{equation}
	g_{y_i y_i} = 1 - \frac{7-q}{8} \frac{r_0^n}{r^n} s^2_{\kappa} u_t^2 - \frac{7-q}{8} s^2_{\kappa} H_{tt}
	 \left[ 1 - \left( 1 + \frac{7-q}{8} \right)\frac{r_0^n}{r^n} s^2_{\kappa} u_t^2 \right] + \mathcal{O}\left(\frac{r_0^{n+2}}{r^{n+2}}\right) ~~.
\end{equation}

\subsubsection*{Dipole terms}

The dipole terms can be read off from the asymptotic expansion of the solution as before and read
\begin{equation}
\begin{split}
	\hat{f}^{(D)}_{tt} = \left( 1 + \frac{7-q}{8} s^2_{\kappa} \right) f^{(D)}_{tt} ~~,~~
	\hat{f}^{(D)}_{t z_i} = c_{\kappa} f^{(D)}_{t z_i} ~~,~~
	\hat{f}^{(D)}_{z_i z_i} = \frac{q+1}{8} s^2_{\kappa} f^{(D)}_{tt} + f^{(D)}_{z_i z_i} ~~,~~
\end{split}
\end{equation}
\begin{equation*}
\begin{split}
	\hat{f}^{(D)}_{y_i y_i} = - \frac{7-q}{8} s^2_{\kappa} f^{(D)}_{tt} ~~,~~
	\hat{f}^{(D)}_{r r} = \frac{q+1}{8} s^2_{\kappa} f^{(D)}_{tt} + f^{(D)}_{rr} ~~,~~
	\hat{f}^{(D)}_{\Omega \Omega} = \frac{q+1}{8} s^2_{\kappa} f^{(D)}_{tt} + f^{(D)}_{\Omega\Omega}~~,
\end{split}
\end{equation*}
where $\hat{f}_{\mu\nu}$ denote the dipole coefficients of the charged solution.

The dipole contribution to the stress-energy tensor \eqref{stpd} can be extracted using the method outlined in Sec.~\ref{sec:solutions}. The transverse gauge condition given by Eq.~\eqref{eqn:transversegaugecond} takes the same form as Eq.~\eqref{eqn:ubrTraceCond1} which ensures again that the relation given in Eq.~\eqref{eqn:ubrTraceCond2} remains the same after the successive T-duality transformations. We can therefore use Eq.~\eqref{eqn:dmunuCoeff} to read off the bending moment and obtaining the only non-zero components
\begin{equation} \label{eqn:dcompqm}
\begin{split}
	\hat{d}_{tt} &= c^2_{\kappa} f^{(D)}_{tt} + f^{(D)}_{\Omega\Omega} ~~,~~
	\hat{d}_{y_i y_j} = - f^{(D)}_{\Omega\Omega} - s^2_{\kappa} f^{(D)}_{tt} ~~, \\
	\hat{d}_{z_i z_j} &= f^{(D)}_{z_i z_j} - f^{(D)}_{\Omega\Omega} ~~,~~
	\hat{d}_{t z_i} = c_{\kappa} f^{(D)}_{t z_i} ~~.
\end{split}
\end{equation}
The non-vanishing dipole terms appearing in the higher-form gauge field expansion read
\begin{equation}
\begin{split}
	a^{(D)}_{t y_1 \ldots y_q } = c_{\kappa}s_{\kappa} f^{(D)}_{tt}  ~~,~~
	a^{(D)}_{z_i y_1 \ldots y_q} = s_{\kappa} f^{(D)}_{t z_i} ~~.
\end{split}
\end{equation}
This class of solutions are special in the sense that the charge is always smeared along the directions along which the brane is bent. In other words, the directions in which the $q$-brane charge lies are always flat and therefore never critically boosted. It is therefore possible to introduce a set of vectors $v_{a}^{(i)}~,~i=1...q$ describing the $q$ directions in which the smeared $q$-charge is located (see Sec.~\ref{sec:physical}). The bending moment given in Eq.~\eqref{eqn:dcompqm} can then be written in terms of the generalized Gibbons-Maeda boost and charge parameters using the relations \eqref{eqn:GMrelations} and read
\begin{equation} \label{eqn:dipolestressqm}
\begin{split}
	\hat{d}_{ab} =\;
	& - \xi_2(n) \cosh^2\alpha \left[ \frac{ K_{ab}}{(n+2)\cosh^2\alpha} + 2 u^c u_{(a}K_{b)c} + \frac{3n+4}{n^2(n+2)} \eta_{ab} K \right]  \\ 
	& - \bar{k} \xi_2(n) \left[n (\cosh^2\alpha \; u_au_b - \sinh^2\alpha \sum_{i=1}^{q} v_a^{(i)}v_b^{(i)}) - \eta_{ab} \right]K   ~~,
\end{split}
\end{equation}
under the same assumptions on $K_{ab}$ as before. Furthermore, the electric dipole moment is obtained using Eq.~\eqref{eqn:dipoleCurrent} and can be written as
\begin{equation} \label{eqn:dipolescurrentqm}
	\hat{p}_{b a_1 \ldots a_q} =\;
	 - (q+1)! \xi_2(n) \cosh\alpha \sinh\alpha \left[ u^c v_{[a_1}^{(1)} ... v_{a_q}^{(q)} K_{b]c} + \bar{k} u_{[b}v_{a_1}^{(1)} ... v_{a_q]}^{(q)} K \right] \,.
\end{equation}
The two dipole moments \eqref{eqn:dipolestressqm} and \eqref{eqn:dipolescurrentqm} have an identical form when compared to the dipole moments given by Eq.~\eqref{eqn:dipolestressq0} and Eq.~\eqref{eqn:dipolescurrentq0}. This is perhaps not too surprising, since the extrinsic perturbations of all the solutions are always along the smeared directions, i.e., the type of bending is similar for all the solutions considered here. From the bending and electric dipole moments one can read off the corresponding response coefficients and these are presented in Sec.~\ref{sec:solutions}.

\addcontentsline{toc}{section}{References}
\footnotesize
\providecommand{\href}[2]{#2}\begingroup\raggedright\endgroup

\end{document}